\documentclass[sigconf, anonymous=False, review=False]{acmart}
\usepackage{multirow}
\usepackage{makecell}
\usepackage{subfig}
\usepackage{bbding}
\usepackage{enumitem}
\usepackage{colortbl}
\usepackage{diagbox}
\usepackage{comment}
\usepackage{amsmath}
\usepackage{algorithm}
\usepackage{algpseudocode}

\AtBeginDocument{%
  }
\usepackage{newfloat}
\usepackage{listings}

\setcopyright{acmlicensed}
\copyrightyear{2018}
\acmYear{2018}
\acmDOI{XXXXXXX.XXXXXXX}
\acmConference[Conference acronym 'XX]{Make sure to enter the correct
  conference title from your rights confirmation email}{June 03--05,
  2018}{Woodstock, NY}

\acmISBN{978-1-4503-XXXX-X/2018/06}

\begin{document}

\title{SPAR: Enhancing Industrial-Scale Generative POI Recommendation via Spatial Perception}


\author{Fangye Wang}
\affiliation{%
  \institution{AMAP, Alibaba Group}
  \city{Beijing}
  \country{China}
}
\email{wangfangye.wfy@alibaba-inc.com}

\author{Yunjin Gu}
\authornote{Work done during the internship at AMAP, Alibaba Group}
\affiliation{
  \institution{The Chinese University of Hong Kong, Shenzhen}
  \city{Shenzhen}
  \country{China}}
\email{yunjingu@link.cuhk.edu.cn}

\author{Haowen Lin}
\affiliation{%
  \institution{AMAP, Alibaba Group}
  \city{Beijing}
  \country{China}
}
\email{linhaowen.lhw@taobao.com}

\author{Yifang Yuan, Song Yang}
\affiliation{%
  \institution{AMAP, Alibaba Group}
  \city{Beijing}
  \country{China}
}
\email{{yuanyifang.yyf, song.yangs}@alibaba-inc.com}



\author{Xiaojiang Zhou}
\authornote{Corresponding Author}
\affiliation{%
  \institution{AMAP, Alibaba Group}
  \city{Beijing}
  \country{China}
}
\email{zhouxiaojiang.zxj@taobao.com}

\author{Pengjie Wang}
\affiliation{%
  \institution{AMAP, Alibaba Group}
  \city{Beijing}
  \country{China}
}
\email{pengjie.wpj@alibaba-inc.com}


\begin{abstract}
Generative Point-of-Interest (POI) recommendation,  which autoregressively generates a target POI's semantic ID (SID), holds great promise for Location-Based Services, where a recommendation helps only if the user can reach it. Yet, existing methods primarily model behavioral preferences through interaction sequences and collaborative signals. Although some incorporate geographic information into SID construction, their subsequent training is not systematically designed to learn and retain urban spatial relations involving distance, direction, and reachability, so behaviorally plausible recommendations may still be geographically unsuitable. We argue that such services require injecting real urban spatial knowledge into the interest space, rather than inferring geography from behavior alone. Hence, we propose SPAR, a unified framework whose three synergistic stages jointly construct, cultivate, and preserve urban spatial knowledge: (1) at the tokenization level, Spatially-Intrinsic SID (SI-SID) explicitly encodes longitude--latitude coordinates into a sinusoidal geospatial embedding and fuses it with the textual semantic embedding, producing identifiers via RQ-Kmeans that are simultaneously semantically and geographically consistent; (2) at the cognition level, Multi-Granular Geospatial CPT (MG-CPT) continually pre-trains the base LLM on 25 curated geospatial datasets organized into three tiers of basic attributes, pairwise relations, and city-scale navigation, so that scattered POIs cohere into a connected urban space; and (3) at the adaptation level, Task-Vector Anchored SFT (TV-SFT) anchors the acquired spatial knowledge as a frozen parameter-space task vector to mitigate catastrophic forgetting of spatial knowledge during behavioral fine-tuning, thereby fusing the two spaces. Extensive quantitative and visualization experiments on two public and four industrial-scale datasets demonstrate the effectiveness of SPAR. 
\end{abstract}



\maketitle
\section{Introduction}
With the pervasiveness of Location-Based Services (LBS), next Point-of-Interest (POI)~\cite{wang2025tool4poi,chen2025onesearch,liu2024nextlocllm,feng2024rotan,zhang2025survey} Recommendation has become a cornerstone capability of LBS platforms such as Kuaishou~\cite{deng2025onerec}, AMAP~\cite{lv2026reasoning} and Google~\cite{he2026plum}. Given a user's historical behavior and real-time context (e.g., query and location), it recommends the POI the user will visit next. Serving billions of users and POIs, its quality directly shapes users' decision-making efficiency and experience, making it a central problem for both academia and industry.

Early studies formulate the next-POI recommendation as a sequential prediction task, evolving from Markov chains and matrix factorization to RNN~\cite{graves2012long, hidasi2015session} and Transformer-based~\cite{luo2023timestamps, li2024llm4poi_rec} models that capture behavioral preferences from historical trajectory. Recently, Generative Recommendation Models (GRMs)~\cite{wang2025gnpr, zhang2025survey} based on a Large Language Model (LLM)~\cite{naveed2025comprehensive,yang2025qwen3} have emerged as a powerful paradigm. Training a GRM typically involves two steps: each POI is first tokenized into a Semantic ID (SID) from its semantic representation via residual quantization (e.g., RQ-VAE~\cite{wang2025gnpr} or RQ-Kmeans~\cite{wei2025oneloc}); the LLM is then supervised fine-tuned (SFT) on user behavior sequences rewritten with these SIDs, enabling it to autoregressively generate the SID of the target POI~\cite{he2026plum,deng2025onerec, wang2025gnpr}. Since SIDs are newly added tokens, some recent methods further insert a continued pre-training (CPT) step beforehand to align the new SID modality with the LLM's existing knowledge~\cite{he2026plum, wang2026geogr}. Revisiting this pipeline, geographic information is incorporated into SID construction in some methods~\cite{wang2025gnpr,chen2026revisiting,he2026birds}, but subsequent CPT and SFT are not systematically designed to learn and retain urban spatial relations. Consequently, explicit geographic grounding and city-specific spatial knowledge remain insufficiently developed alongside behavioral preference modeling. 


\textbf{Imprecise geographic encoding at the tokenization stage.} Longitude--latitude coordinates are either serialized as plain textual semantics, whose digit tokens carry no metric structure, or exploited only indirectly through collaborative signals~\cite{wang2025gnpr,liu2025onerec_think} and contrastive objectives~\cite{wang2026geogr, chen2026revisiting}. Such signals are themselves largely derived from co-occurrence and thus cannot substitute for explicit geographic measurement. Quantization is therefore dominated by high-variance semantic features while geography degenerates into a weak regularizer, and the resulting identifier space lacks spatial continuity: POIs with adjacent coordinates are not guaranteed adjacent SIDs, while semantically similar yet distant POIs (e.g., two branches of the same chain located tens of kilometers apart in the same city) may share near-identical ones. An effective SID should encode not only what a POI is but also where it lies in the city.

\textbf{Absent urban spatial knowledge at the training stage.} In LBS, existing POI-related GRMs~\cite{wang2026geogr,wang2025gnpr} still rely mainly on SFT to model sequences of user behavior, with CPT, when used at all, not aimed at urban spatial knowledge. This overlooks what makes recommendation in such services distinctive: next-POI prediction is essentially movement through real urban space. Users are not merely reproducing co-occurrence patterns but physically traveling, so beyond straight-line proximity they weigh direction and reachability. However, a general LLM's web-scale pre-training offers little fine-grained knowledge of a specific city, and existing CPT, typically a small-scale warm-up over POI textual attributes, does little to supply it, leaving the road network and its connectivity unlearned. Optimized on behavior alone, SFT captures only the interest side: the model learns which POIs are frequently visited together, but not whether a candidate is accessible from the user's current location—a POI two kilometers away in a straight line may lie across a river or require a lengthy detour. Its predictions are thus behaviorally plausible, yet ungrounded in real-world geography. Worse still, even injecting spatial knowledge with CPT is not enough: under the massive behavioral data, standard SFT tends to overwrite it, so unless the knowledge is explicitly preserved during adaptation, the model drifts back to behavior-only fitting.

To address these deficiencies, we propose \textbf{SPAR}, a unified framework that injects real urban spatial knowledge into the interest space of generative POI recommendation, rather than inferring geography from behavior alone. SPAR realizes this through three synergistic stages that respectively construct, cultivate, and preserve this knowledge. First, at the representation level, \textbf{Spatially-Intrinsic SID (SI-SID)} uses an explicit sinusoidal encoder to lift raw longitude–latitude coordinates into a geospatial embedding, fuses it with the POI's semantic embedding, and discretizes the result via RQ-Kmeans, so that every identifier is anchored to a concrete location. Second, at the cognition level, \textbf{Multi-Granular Geospatial CPT (MG-CPT)} continually pre-trains the backbone LLM on 25 datasets that span three tiers of increasing spatial complexity and together comprise millions of instances built from POI, road, and navigation data, internalizing the spatial relations of the whole city. Finally, at the adaptation level, \textbf{Task-Vector Anchored SFT (TV-SFT)} adapts the model to user behavior while retaining the CPT-induced parameter update as a frozen task vector, with the aim of mitigating spatial knowledge forgetting during fine-tuning. SPAR thereby combines behavioral preference modeling with learned urban spatial relations when generating the target POI's SID. The major contributions are summarized as follows:


%


\begin{itemize}[leftmargin=1em]
    \item We propose SPAR, a unified framework that injects real urban spatial knowledge into the interest space of generative POI recommendation; its three synergistic stages jointly construct, cultivate, and preserve this knowledge.
    \item SI-SID fuses geospatial and semantic embeddings before residual quantization, making geographic topology an intrinsic property of the identifier codebook.
    \item MG-CPT continually pre-trains the LLM on 25 multi-tier POI, road, and navigation datasets, internalizing the city's spatial relations so that POIs become places related by distance, direction, and reachability. TV-SFT then anchors this knowledge as a task vector, grounding behavioral fitting in urban spatial relations.
    \item Comprehensive experiments on two public benchmarks and four industrial-scale datasets demonstrate SPAR's superior recommendation performance and robustness under sparse behavioral histories. Ablation studies, spatial cognition evaluations, and qualitative analyses further support the complementary contributions of SI-SID, MG-CPT, and TV-SFT to spatially informed POI recommendation.
\end{itemize}

\section{Related Work}
\textbf{POI Recommendation.}
Traditional next-POI recommendation methods are typically formulated as a sequential prediction task, where the goal is to predict a user's next visited location based on their historical check-in sequence~\cite{zhai2025cognitive,li2024llm4poi_rec,wang2025gnpr,kang2018self,sun2019bert4rec}. With the advent of deep learning, recurrent neural networks (RNNs)—particularly LSTM and GRU variants—were widely adopted to capture sequential preferences~\cite{hidasi2015session}, followed by transformer-based models that demonstrated superior performance through self-attention mechanisms~\cite{kang2018self}. For instance, STAN~\cite{luo2021stan} explicitly models pairwise spatio-temporal relationships via interpolated distance encoding, while GeoSAN~\cite{lian2020geography} incorporates geographical influence into self-attention for next-POI recommendation. More recently, LLMs have been integrated to leverage commonsense knowledge for semantic-aware prediction~\cite{li2024llm4poi_rec, wang2025gnpr}. In addition, graph neural networks gained traction for modeling user-POI interactions~\cite{luo2023timestamps,yang2022getnext}.

\textbf{Generative Recommendation.}
Beyond discriminative models, LLM-based GRMs~\cite{liu2024end,wei2025oneloc,zhang2025survey} have emerged as a powerful paradigm that directly generates item identifiers, eliminating the multi-stage cascaded pipeline of retrieval and ranking. TIGER~\cite{rajput2023recommender} pioneers this direction with hierarchical SIDs learned via RQ-VAE~\cite{lee2022autoregressive}, while subsequent works adopt RQ-Kmeans for more stable quantization~\cite{deng2025onerec}. In industrial practice, OneRec~\cite{deng2025onerec} and OneRec-V2~\cite{zhou2025onerec} unify retrieval and ranking into end-to-end generative frameworks with reinforcement learning. OneLoc~\cite{wei2025oneloc} incorporates geographic information into generative recommendation for local lifestyle services with reinforcement learning. GeoGR~\cite{wang2026geogr} and GNPR-SID~\cite{wang2025gnpr} refine SIDs with contrastive learning to better capture POI semantics and collaborative signals. PLUM~\cite{he2026plum} employs an LLM to extract features and represent items as SIDs for generative recommendation for YouTube recommendation with CPT and SFT~\cite{gu2026deep_survey}. Nevertheless, incorporating geographic information into identifiers or behavioral training does not by itself establish systematic learning and retention of urban spatial relations. SPAR addresses this gap by integrating explicit geographic grounding, geospatial continued pre-training, and task-vector anchored adaptation into generative POI recommendation.


\section{Problem Formulation}
\label{sec:problem}

Let $\mathcal{U} = \{ u_1, u_2, ..., u_{M} \}$ and $\mathcal{P} = \{ p_1, p_2, ..., p_N \}$ denote the set of $M$ users and $N$ POIs, respectively. Each POI is defined as a tuple $p_i = (\text{lon}_i, \text{lat}_i, \text{addr}_i, \text{ctx}_i)$, capturing geographical coordinates, address, and contextual attributes (e.g., brand, category). Meanwhile, each POI $p_i$ is associated with a SID $s_i$, generated via the RQ-Kmeans method. For each user $u \in \mathcal{U}$, we record their interaction trajectory (i.e., interaction history) as $T^u = \{ r^u_1, r^u_2, ..., r^u_n \}$, ordered in reverse chronological order. Each interaction record $r^u_t = (u, p, t, \text{con}, a)$ indicates that user $u$ interacted with POI $p$ at timestamp $t$ under condition $\text{con}$ with action $a$. In the AMAP application scenario, action $a$ encompasses behaviors such as navigation, reservation, booking, and collection, while condition $\text{con}$ captures real-time context, such as query, current coordinates, and other metadata. 


\textbf{Next POI Recommendation.}  Given a user $u$'s historical trajectory $T^u$ and a real-time condition $\text{con}_u$, the task is to predict the next POI $p_{n+1}$ (i.e., its SID $s_{n+1}$) by maximizing the conditional probability: $p_{n+1}^* = \arg\max_{p \in \mathcal{P}} \mathbb{P}(p | u, T^u, \text{con}_u)$. Unlike traditional discriminative methods that score all candidates, SPAR formulates this as a generative task~\cite{wang2025gnpr} where $s_{n+1}$ is autoregressively decoded from the sequence $(s_1, s_2, ..., s_n)$ conditioned on $\text{con}_u$.

\label{sec:sisid}
\begin{figure*}[t]
    \centering
    \includegraphics[width=0.99\textwidth]{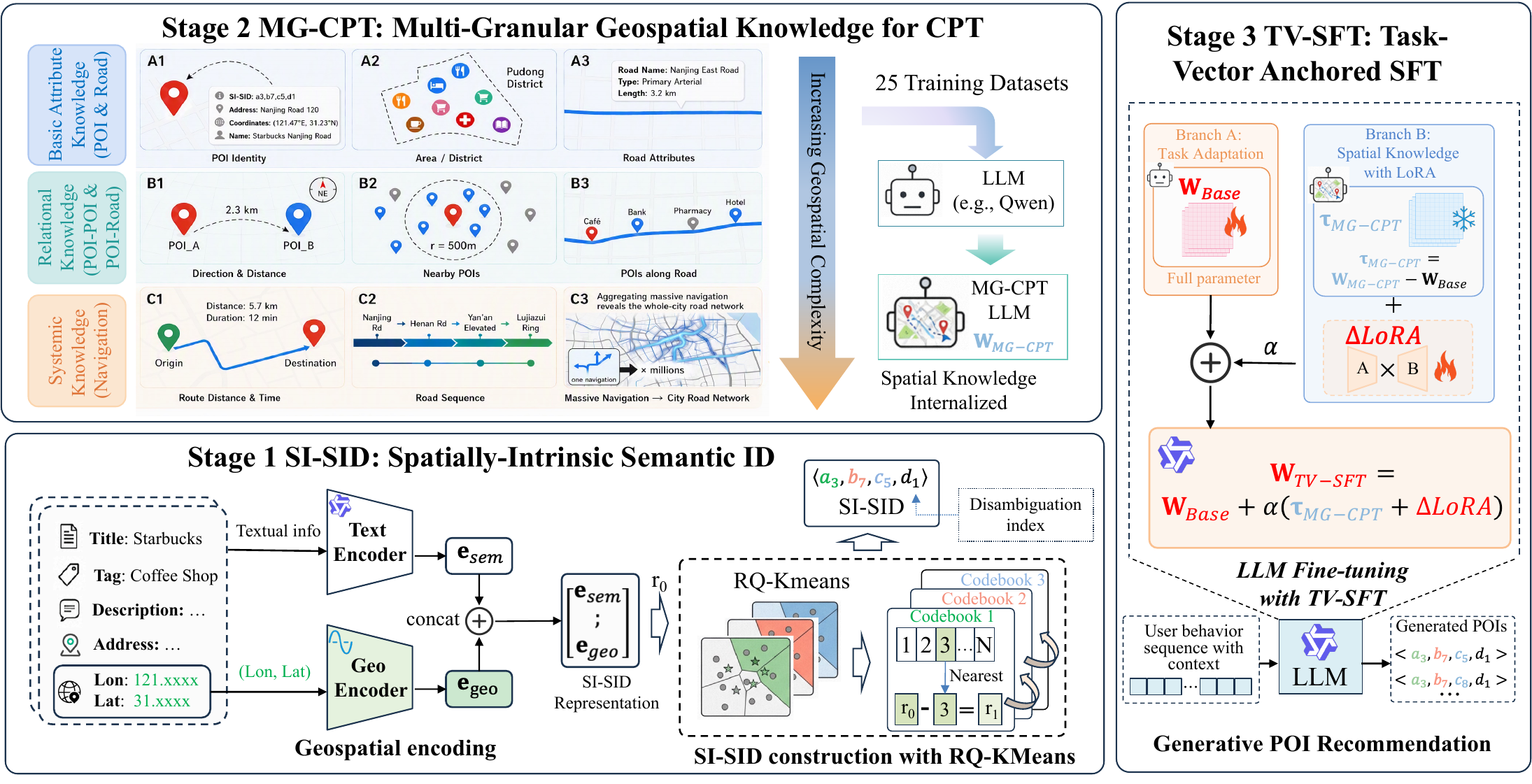}
    \caption{\textbf{SPAR} enhances large-scale generative POI recommendation with urban spatial knowledge. Stage 1: \textbf{SI-SID} constructs a Spatially-Intrinsic SID by fusing textual ($\mathbf{e}_{sem}$) and geospatial ($\mathbf{e}_{geo}$) embeddings and quantizing them via RQ-Kmeans. Stage 2: \textbf{MG-CPT} continually pre-trains the LLM on multi-granular geospatial datasets spanning basic-attribute, relational, and systemic spatial knowledge. Stage 3: \textbf{TV-SFT} jointly performs full-parameter task adaptation ($\mathbf{W}_{\text{Base}}$) and anchored spatial-knowledge refinement, where the frozen $\tau_{\text{MG-CPT}}$ is adjusted by a trainable $\Delta{\text{LoRA}}$, giving $\mathbf{W}_{\text{TV-SFT}}$ for POI recommendation.}
    \label{fig:spar_overview}
\end{figure*}

\section{Methodology: SPAR}

We introduce SPAR, a generative POI recommendation framework with three stages (Figure~\ref{fig:spar_overview}): SI-SID constructs the identifier space, fusing a POI's semantic embedding with a geospatial encoding of coordinates and quantizing them into spatial semantic tokens. MG-CPT cultivates spatial reasoning, continually pre-training the LLM on multi-granular geospatial data so scattered POIs cohere into a connected urban space. TV-SFT preserves this knowledge, anchoring it as a frozen task vector while the model is fine-tuned on user behavior, fusing the interest space of user trajectories with the real urban space to generate the target POI's SI-SID.

\subsection{Spatially-Intrinsic SID (SI-SID)}

Recent GRMs derive an item's SID by encoding its textual descriptions into a semantic embedding via an LLM and then discretizing it through RQ-VAE or RQ-Kmeans. In LBS, however, a user's interactions are tightly coupled with real-time geographic context, as users tend to choose POIs closer to their current location, which such semantics-driven identifiers fail to reflect. We therefore design SI-SID (Figure~\ref{fig:spar_overview} stage 1), which incorporates geographic coordinates into the discrete token sequence itself, realized in two steps: explicit geospatial encoding and SI-SID construction with RQ-Kmeans.

\subsubsection{Explicit Geospatial Encoding} 
\label{sec:geo_encoding}  

Raw longitude--latitude coordinates provide a low-dimensional representation of location but do not directly expose spatial relationships to the model. We therefore adopt an explicit sinusoidal geospatial encoder inspired by NeRF~\cite{mildenhall2021nerf,russwurm2024geographic} and Transformer positional encodings~\cite{vaswani2017attention,naveed2025comprehensive}. Within the city-scale setting considered here, given a POI with longitude $\lambda$ and latitude $\phi$, we first normalize each coordinate to $[0,2\pi]$:
\begin{equation}     \hat{\lambda} = \frac{\lambda + 180}{360} \cdot 2\pi, \qquad     \hat{\phi} = \frac{\phi + 90}{180} \cdot 2\pi. 
\end{equation} 
Each is then projected onto $K$ exponentially spaced frequency bands. Hyperparameter $K$ controls the granularity of the encoding: 
\begin{equation}     f_k = 10000^{-k/K}, \qquad k = 0, \dots, K-1, \end{equation} 
yielding the Fourier geospatial features $\mathbf{\gamma}(\hat{\lambda}) \in \mathbb{R}^{2K} $: 
\begin{equation}     \gamma(\hat{\lambda}) = \left[\sin(f_0 \hat{\lambda}), \cos(f_0     \hat{\lambda}), \dots, \sin(f_{K-1}\hat{\lambda}),     \cos(f_{K-1}\hat{\lambda})\right]  
\end{equation}  
and analogously $\gamma(\hat{\phi})$. The final geospatial representation $\mathbf{e}_{\text{geo}} \in \mathbb{R}^{d_g}  $ is obtained by concatenating $\gamma(\hat{\lambda})$, $\gamma(\hat{\phi})$, and $d_g=4K$.


\textbf{Spatial properties of the encoding.}
Let $u \in [0, 2\pi]$ denote either normalized coordinate $\hat{\lambda}$ or $\hat{\phi}$; the following properties hold for each coordinate separately and extend to their concatenation.
\textbf{Property 1 (Shift-invariant kernel).} By the product-to-sum identity, $\langle \gamma(u_1), \gamma(u_2) \rangle = \sum_{k=0}^{K-1}\big[\sin(f_k u_1)\sin(f_k u_2) + \cos(f_k u_1)\cos(f_k u_2)\big] = \sum_{k=0}^{K-1} \cos\big(f_k (u_1 - u_2)\big)$ depends only on the displacement $u_1 - u_2$ rather than on absolute positions; since $f_k \le f_0 = 1$, every term is decreasing in $|u_1 - u_2|$ on $[0, \pi]$, a range that covers all intra-city displacements by a wide margin, so embedding similarity decreases monotonically with geographic displacement.
\textbf{Property 2 (Lipschitz continuity).} Applying $1 - \cos\theta \le \theta^2/2$ to the squared distance gives $\| \gamma(u_1) - \gamma(u_2) \|_2^2 = \sum_{k=0}^{K-1} 2(1 - \cos(f_k (u_1 - u_2))) \le (\sum_{k=0}^{K-1} f_k^2)(u_1 - u_2)^2$, so the encoding is Lipschitz-continuous.
POIs with adjacent coordinates thus obtain adjacent geospatial embeddings $\mathbf{e}_{\text{geo}}$ and, after the residual quantization described next, tend to share SI-SID prefixes, making geographic proximity an intrinsic property of the identifier space. We refer readers to the supplementary material for a more detailed analysis of these properties and their empirical verification.

\subsubsection{SI-SID Construction with RQ-Kmeans}
\label{sec:rq_kmeans}

We generate the POI's semantic embedding
$\mathbf{e}_{sem} \in \mathbb{R}^{d_s}$ from a pretrained text encoder
(e.g., Qwen3-embedding-4B~\cite{yang2025qwen3}) and fuse it with the geospatial embedding via concatenation followed by L2 normalization:
\begin{equation}
    \mathbf{x} = [\mathbf{e}_{sem};\, \mathbf{e}_{geo}] \,/\, \| [\mathbf{e}_{sem};\, \mathbf{e}_{geo}] \|_2 \in \mathbb{R}^{d_s + d_g}.
\end{equation}
This normalization places the fused representation on the unit hypersphere before residual quantization. 

The fused embedding is then discretized by an $L$-layer residual K-means quantizer. Let $\mathbf{r}^{(0)} = \mathbf{x}$ for each POI. Each layer's codebook $\mathcal{C}_\ell = \{c_\ell^k\}_{k=1}^{N_c}$ is constructed by applying K-means clustering to the residuals collected from all training POIs at the previous layer:
\begin{equation}
    \mathcal{C}_\ell = \mathrm{K\text{-}Means}(R_\ell,\, N_c),
    \qquad
    R_\ell = \{\mathbf{r}_i^{(\ell-1)}\}_{i=1}^{N},
\end{equation} 
where $N_c$ is the codebook size ($N_c = 512$ for industrial-scale datasets). Given the trained codebooks, each POI is quantized layer by layer:
\begin{align}
    c_\ell^* &= \arg\min_{c \in \mathcal{C}_\ell} \| \mathbf{r}^{(\ell-1)} - c \|_2, \\
    \mathbf{r}^{(\ell)} &= (\mathbf{r}^{(\ell-1)} - c_\ell^*) \,/\, (\| \mathbf{r}^{(\ell-1)} - c_\ell^* \|_2 + \epsilon).
\end{align}
Here, $\epsilon=10^{-8}$ prevents division by zero. The resulting SI-SID has $L=4$ levels: three codewords obtained through residual K-means quantization and a final disambiguation index that distinguishes POIs sharing the first three codewords~\cite{wang2025gnpr}. We represent a POI's SI-SID as $(a_*,b_*,c_*,d_*)$, for example, $\langle a_3,b_4,c_1,d_1\rangle$. This identifier provides the spatially grounded representation on which MG-CPT learns urban spatial relations and TV-SFT performs downstream adaptation.

\subsection{Multi-Granular Geospatial CPT (MG-CPT)}
\label{sec:mg_cpt}

While SI-SID provides a structurally sound carrier for spatial perception, the base foundation model still cannot reason over the geographic relationships encoded in these identifiers. We therefore propose Multi-Granular Geospatial CPT (MG-CPT), which continually pre-trains a foundation LLM on real-world POI, road, and navigation data organized into three tiers of increasing spatial complexity. This is motivated by real LBS behavior: users choose destinations by reasoning about distance, direction, and reachability, and MG-CPT explicitly instills these forms of spatial knowledge. By jointly learning from all three tiers, MG-CPT internalizes the spatial relations of the whole city into the model's parameters, so that each SI-SID denotes not an isolated POI but a place connected to others through distance, direction, and reachability.

\subsubsection{Three-Tier Geospatial Data Taxonomy}
As illustrated in Figure~\ref{fig:spar_overview} stage 2, these tiers are instantiated as training datasets, of which the figure depicts only a few representative ones. In total, we construct 25 training datasets that jointly characterize the real-world profile of POIs, their relationships with the surrounding road network, and a dynamic depiction of the city's road system by aggregating large-scale navigation data, together comprising millions of training instances. The three tiers are summarized below, with full details deferred to the supplementary material.

\textbf{Tier 1: Basic Attribute Knowledge (POI \& Road).} This tier grounds each entity in its intrinsic attributes, linking a POI's SI-SID to its name, category, address, district, and coordinates, and a road to attributes such as name, type, and length, so that the learned SI-SID tokens denote concrete geographic entities rather than abstract symbols.

\textbf{Tier 2: Relational Knowledge (POI-POI \& POI-Road).} This tier teaches pairwise spatial relations of two complementary kinds. One is \textit{real-world aggregation}: the POIs surrounding a location and those lying along a road, reflecting how places actually cluster in the city. The other is \textit{metric reasoning}: the direction and Euclidean distance between POIs, adding the orientation and proximity computations users rely on when comparing candidate destinations.

\textbf{Tier 3: Systemic Knowledge (Navigation).} Euclidean distance alone can be misleading: two POIs close in straight-line distance may be hard to reach due to rivers, overpasses, or one-way roads. To capture real reachability rather than mere proximity, this tier explicitly introduces navigation data in multiple forms (route distance, travel time, and ordered road sequences), letting the model internalize the connectivity of the whole-city road network instead of memorizing individual routes.

\subsubsection{Continued Pre-Training}
All datasets are serialized into next-token-prediction instances and jointly used to continually pre-train a base LLM under the standard causal language modeling objective:
\begin{equation}
    \mathcal{L}_{\text{MG-CPT}} = -\sum\nolimits_{t} \log P(x_{t+1} \mid x_1, \dots, x_t),
\end{equation}
where $x_{1:t}$ denotes a training sequence from the mixed corpus. This process extends the base LLM, with parameters $\mathbf{W}_{\text{Base}}$, into an urban-space-aware LLM with parameters $\mathbf{W}_{\text{MG-CPT}}$. As evaluated on an 18-task benchmark in Section~\ref{sec:mgcpt_exp}, the trained MG-CPT LLM exhibits substantially enhanced knowledge of urban space, providing a robust foundation for the subsequent TV-SFT stage.

\subsection{Task-Vector Anchored SFT (TV-SFT)}
\label{sec:tv_sft}

MG-CPT provides a spatially informed foundation for next-POI recommendation, allowing behavioral trajectories to be modeled together with urban spatial relations rather than through behavioral transitions alone. Downstream adaptation nevertheless requires fine-tuning on large-scale user-behavior data, which may weaken the spatial capabilities acquired during CPT. To retain a fixed CPT-induced parameter component while allowing task-specific adaptation, we draw inspiration from model editing and knowledge injection~\cite{yadav2023ties,ilharco2022editing,liu2025pave}, which represent task-specific parameter updates as directions in parameter space. We propose TV-SFT, combining a trainable base branch with a frozen CPT task vector and a trainable low-rank correction (Figure~\ref{fig:spar_overview}, Stage 3). We first define the MG-CPT task vector as:
\begin{equation}
    \tau_{\text{MG-CPT}} = \mathbf{W}_{MG-CPT} - \mathbf{W}_{base}.
\end{equation}

The task vector records the parameter update induced by MG-CPT and provides a fixed reference for downstream adaptation. TV-SFT combines two components. \textbf{(i) Task Adaptation} performs full-parameter SFT through the base branch $\mathbf{W}_{\text{Base}}$ to model user behavior. \textbf{(ii) Spatial Anchoring} retains $\tau_{\text{MG-CPT}}$ as a frozen parameter component and learns a low-rank correction $\Delta\text{LoRA}$. For each adapted linear layer, the correction takes the form $\mathbf{BA}$, where $\mathbf{B}\in\mathbb{R}^{d_{\text{out}}\times r}$, $\mathbf{A}\in\mathbb{R}^{r\times d_{\text{in}}}$, and $r$ is the rank hyperparameter. Only $\mathbf{W}_{\text{Base}}$ and the LoRA parameters are trainable, while $\tau_{\text{MG-CPT}}$ remains fixed. This formulation explicitly retains the CPT-induced parameter component while allowing task-specific adaptation through the trainable base branch and low-rank correction.


The two branches are combined into the final fine-tuned model:
\begin{equation}
    \mathbf{W}_{\text{TV-SFT}} = \mathbf{W}_{\text{Base}} + \alpha\,(\tau_{\text{MG-CPT}} + \Delta\text{LoRA}).
\end{equation}
Hyperparameter $\alpha$ scales the frozen task vector together with the low-rank correction and is set to $1$ by default. Let $\mathbf{W}_{\text{Base}}^{(0)}$ denote the base-branch initialization. By construction, $\mathbf{W}_{\text{Base}}^{(0)}+\tau_{\text{MG-CPT}}=\mathbf{W}_{\text{MG-CPT}}$. If $\Delta\text{LoRA}$ is initialized to zero, $\alpha=1$ initializes the effective model at the MG-CPT checkpoint, whereas $\alpha=0$ removes the anchored component and yields full-parameter SFT of the base model. During training, $\mathbf{W}_{\text{Base}}$ changes while $\tau_{\text{MG-CPT}}$ remains fixed.

\textbf{Training objective.} Given a prompt composed of the user's historical behavior sequence and contextual signals, the fine-tuned LLM learns to autoregressively generate the SI-SID of the target POI $\mathbf{Q}_t = \langle a_*, b_*, c_*, d_*\rangle $ by minimizing the negative log-likelihood:
\begin{equation}
    \begin{aligned}
    \mathcal{L}_{\text{TV-SFT}}
    &= -\log P_{\mathbf{W}_{\text{TV-SFT}}}(\mathbf{Q}_t \mid \text{prompt}) \\
    &= -\sum\nolimits_{i=1}^{4} \log P_{\mathbf{W}_{\text{TV-SFT}}}(q_i \mid \text{prompt},\, q_{<i}),
    \end{aligned}
\end{equation}
where $q_i$ denotes the $i$-th token of $\mathbf{Q}_t$ and $q_{<i}$ the preceding tokens.

\section{Experiments}
\subsection{Experimental Setting}



\begin{table}[t]
    \centering
    \caption{Statistics of the processed datasets.}
    \label{tab:dataset_stats}
    \setlength{\tabcolsep}{2.6mm}
    \scalebox{0.83}{
    \begin{tabular}{cccccc}
    \hline
    \hline
    Type & Dataset & \#Users & \#POIs & \#Instances & Sparsity \\
    \hline
    \multirow{2}{*}{Public} & NYC & 2,083 & 5,135 & 104,074 & 98.10\% \\ 
     & TKY & 2,293 & 7,873 & 361,430 & 98.98\% \\
    \cline{1-6}
    \multirow{4}{*}{Industry} & Beijing & 5,431,217 & 678,819 & 20,980,626 & 99.99\% \\ 
     & Shanghai & 4,819,962 & 919,594 & 22,333,204 & 99.99\% \\ 
     & Tianjin & 2,552,323 & 499,972 & 15,182,776 & 99.99\% \\ 
     & Zhejiang & 11,252,453 & 3,281,512 & 28,068,268 & 99.99\% \\ 
    \hline
    \hline
    \end{tabular}
    }
\end{table}

\textbf{Datasets.} We evaluate SPAR on \textbf{two public benchmarks}, NYC and TKY~\cite{yang2014modeling, wang2025gnpr}, and \textbf{four industrial-scale datasets} collected from AMAP platform, a prominent navigation and mapping platform serving billions of users and POIs. The public datasets are pre-processed following previous works~\cite{wang2025gnpr}: after sorting interactions chronologically, we allocate 80\% for training, 10\% for validation, and 10\% for testing. For industrial dataset, we collect and clean one month of user trajectories and interaction logs. Every interaction therein is treated as one prediction target, for which we trace back the user's preceding behaviors to construct its historical sequence, so that each (history, target) pair forms one complete training instance. The interactions on the last day are reserved as the test set and all earlier ones are used for training. Dataset statistics are shown in Table~\ref{tab:dataset_stats}. We will release the four industrial datasets.

\textbf{Evaluation Metrics.} We evaluate Recall@K and NDCG@K with $K\in\{5,10,20\}$. Recall@K measures whether the target POI appears among the top-K recommendations, while NDCG@K additionally accounts for its rank. We abbreviate these metrics as R@K and N@K in tables where space is limited.

\textbf{Comparison Methods.} We compare SPAR against three categories of baselines: (1) \textit{Traditional sequential models}: SASRec~\cite{kang2018self}, BERT4Rec~\cite{sun2019bert4rec}, GRU4Rec~\cite{hidasi2015session}, Caser~\cite{tang2018caser}, and $S^3$-Rec~\cite{zhou2020s3}; (2) \textit{Transformer based POI recommendation methods}: TPG~\cite{luo2023timestamps} and Rotan~\cite{feng2024rotan}; (3) \textit{Generative recommendation models (GRMs)}: TIGER~\cite{rajput2023recommender}, GNPR-SID~\cite{wang2025gnpr} and PLUM~\cite{he2026plum}.

\begin{table*}[t]
\centering
\caption{Offline performance on NYC and TKY. "Rela.Imp" indicates the relative improvement compared to the best baselines.}
\label{exp:overall_public}
\setlength{\tabcolsep}{2.2mm}
\scalebox{0.81}{
    \begin{tabular}{c|c|ccccc|cc|ccc|cc} 
    \hline
    \hline
    \multicolumn{1}{c|}{\multirow{2}{*}{Dataset}} & \multirow{2}{*}{Metric} & \multicolumn{5}{c|}{Traditional sequential models} & \multicolumn{2}{c|}{POI related} & \multicolumn{3}{c|}{GRMs}&  \multicolumn{1}{c}{\multirow{2}{*}{\makecell{SPAR-4B\\(Rela.Imp)}}} &\multicolumn{1}{c}{\multirow{2}{*}{\makecell{SPAR-8B\\(Rela.Imp)}}}   \\ 
    \cline{3-12}
    \multicolumn{1}{c|}{}  &  & \multicolumn{1}{l}{SASRec} & \multicolumn{1}{c}{BERT4Rec} & \multicolumn{1}{l}{GRU4Rec} & \multicolumn{1}{c}{Caser} & \multicolumn{1}{c|}{$S^3$-Rec} & \multicolumn{1}{l}{TPG} & \multicolumn{1}{c|}{Rotan} & \multicolumn{1}{c}{TIGER} & \multicolumn{1}{c}{GNPR-SID} &\multicolumn{1}{c|}{PLUM}  & \multicolumn{1}{c}{} & \multicolumn{1}{c}{} \\ 
    \hline
    \multirow{6}{*}{NYC}   
& R@5 & 0.3151 & 0.2857& 0.1977  & 0.2883& 0.3071& 0.3551   & 0.4448 & 0.4965 & 0.5311  & \underline{0.5619}&\textit{0.6213 (10.57\%	)}  &\textbf{0.6556 (16.68\%)} \\
& R@10& 0.3896 & 0.3564& 0.2460  & 0.3570& 0.3854& 0.4441   & 0.5223 & 0.5514& 0.5942 & \underline{0.6143}&\textit{0.6686 (8.84\%	)}&\textbf{0.6996 (13.89\%)}  \\
& R@20& 0.4506 & 0.4130& 0.2889  & 0.4135& 0.4503& 0.5121   & 0.5834 & 0.6001& 0.6455  & \underline{0.6642}&\textit{0.7120 (7.20\%)}&\textbf{0.7649 (15.16\%)}  \\
& N@5   & 0.2224 & 0.2074& 0.1442  & 0.2044& 0.2235& 0.2464   & 0.3471 & 0.4131& 0.4430 &\underline{0.4672} &\textit{0.5116 (9.50\%	)}	&\textbf{0.5415 (15.90\%) } \\
& N@10  & 0.2467 & 0.2304& 0.1599  & 0.2267& 0.2489& 0.2755   & 0.3723 & 0.4276& 0.4634  &\underline{0.4870} &\textit{0.5363 (10.12\%	)}&\textbf{0.5651 (16.04\%) } \\
& N@20  & 0.2622 & 0.2448& 0.1708  & 0.2410& 0.2654& 0.2927   & 0.3878 & 0.4443& 0.4766 &\underline{0.4951} &\textit{0.5608 (13.27\%)}&\textbf{0.5906 (19.29\%) } \\
    \hline
    \multirow{6}{*}{TKY}   
& R@5 & 0.3450 & 0.2649& 0.2514  & 0.3257& 0.3365& 0.3725   & 0.4333 & 0.5031& 0.5354 &\underline{0.5526}&\textit{0.6108 (10.53\%	)}&\textbf{0.6474 (17.16\%)}	\\
& R@10& 0.4284 & 0.3326& 0.3106  & 0.4067& 0.4115& 0.4601   & 0.5113 & 0.5808& 0.6130  &\underline{0.6255} &\textit{0.6712 (7.31\%)}&\textbf{0.7098 (13.48\%)}  \\
& R@20& 0.4976 & 0.3943& 0.3651  & 0.4758& 0.4739& 0.5291   & 0.5894 & 0.6431& 0.6675  &\underline{0.6812} &\textit{0.7217 (5.95\%	)}&\textbf{0.7623 (11.91\%)}  \\
& N@5   & 0.2384 & 0.1907& 0.1833  & 0.2273& 0.2423& 0.2591   & 0.3293 & 0.4003& 0.4437 &\underline{0.4659} &\textit{0.4994 (7.19\%)}&\textbf{0.5342 (14.66\%)}	  \\
& N@10  & 0.2655 & 0.2127& 0.2025  & 0.2535& 0.2666& 0.2881   & 0.3568 & 0.4251& 0.4623 &\underline{0.4843}&\textit{0.5116 (5.64\%)} &\textbf{0.5473 (13.01\%)}	  \\
& N@20  & 0.2831 & 0.2284& 0.2163  & 0.2711& 0.2825& 0.3051   & 0.3739 & 0.4401& 0.4788 &\underline{0.4936}&\textit{0.5331 (8.00\%	)} &\textbf{0.5712 (15.72\%)} \\
    \hline
    \hline
    \end{tabular}
}
\end{table*}

\begin{table*}[t]
    \centering
    \caption{Performance comparison on four industrial AMAP Datasets. Ave.Imp denotes the average improvement over PLUM.}
    \setlength{\tabcolsep}{2.5mm}
    \label{exp:private}
    \scalebox{0.94}{
 \begin{tabular}{c|cc|cc|cc|cc|c} 
\hline\hline
Dataset     & \multicolumn{2}{c|}{Beijing}      & \multicolumn{2}{c|}{Shanghai}     & \multicolumn{2}{c|}{Tianjin}      & \multicolumn{2}{c|}{Zhejiang}     &  \\ 
\hline
Model       & Recall@5  & NDCG@5 & Recall@5  & NDCG@5& Recall@5  & NDCG@5& Recall@5  & NDCG@5  & Ave.Imp\\ 
\hline
TIGER (8B)  & 0.4112& 0.2959& 0.3426& 0.2328& 0.3362& 0.2373& 0.4358& 0.2808& -12.31\%\\
GNPR-SID (8B)     & 0.4243& 0.3272& 0.3593& 0.2501& 0.3624& 0.2590& 0.4404& 0.3209& -6.06\%\\
PLUM (8B)   & 0.4514& 0.3403& 0.3704& 0.2712& 0.3910& 0.2761& 0.4665& 0.3518& (Base) \\ 
\hline
SPAR-0.6B   & 0.4729& 0.3546& 0.4062& 0.2975& 0.4310& 0.3083& 0.4892& 0.3618& 7.24\% \\
SPAR-4B     & \underline{0.5708}  & \underline{0.4325}  & \underline{0.4910}  & \underline{0.3656}  & \underline{0.5051}  & \underline{0.3775}  & \underline{0.5424}  & \underline{0.4079}  & \underline{27.38\%}   \\
SPAR-8B     & \textbf{0.6176} & \textbf{0.4643} & \textbf{0.5353} & \textbf{0.3968} & \textbf{0.5528} & \textbf{0.4122} & \textbf{0.5708} & \textbf{0.4266} & \textbf{37.30\%}  \\
\textbf{R.I (SPAR 8B to 4B)} & \textbf{8.20\%} & \textbf{7.35\%} & \textbf{9.02\%} & \textbf{8.53\%} & \textbf{9.44\%} & \textbf{9.19\%} & \textbf{5.23\%} & \textbf{4.58\%} & -\\
 \hline
 \hline
 \end{tabular}
    }
\end{table*}

\textbf{Implementation Details.} All experiments are conducted on 64 NVIDIA A100 GPUs. We use Qwen3-0.6B, -4B, and -8B as the backbone LLMs for MG-CPT and TV-SFT, with their corresponding embedding models producing POI representations. Unless otherwise stated, we use the 4B backbone by default for a balance of effectiveness and efficiency. The SI-SID module uses $L = 4$ levels (three RQ-Kmeans codebooks and one disambiguation index), with $N_c = 32$ for NYC and TKY and $N_c = 512$ for the AMAP datasets. For a fair comparison, all GRMs adopt four-level SIDs, whose last level serves as a disambiguation index. All models are trained with a batch size of 32 and a learning rate of $2\times10^{-5}$.

\subsection{Overall Performance}
\subsubsection{Results on Public Benchmarks.}
As shown in Table~\ref{exp:overall_public}, SPAR achieves state-of-the-art(SOTA) performance across all metrics on NYC and TKY. Two critical observations emerge:

\textbf{GRMs consistently outperform other methods.} Across all six metrics, GRMs (TIGER, GNPR-SID, PLUM, SPAR) consistently surpass traditional and POI-related methods: on NYC R@5, the best GRM (0.6556) exceeds the best traditional one (0.3151) by 108.1\% and the best POI-related one (0.4448) by 47.4\%. Through autoregressive decoding, the generative paradigm more comprehensively models user--POI dependencies, avoiding the representation bottleneck of the candidate-scoring stage in discriminative approaches.

\textbf{SPAR substantially outperforms existing generative baselines.} SPAR-8B surpasses the strongest baseline PLUM (8B) across all metrics, with relative improvements of 11.91\%--19.29\%, confirming the efficacy of reinforcing spatial perception throughout the training lifecycle: SI-SID grounds identifiers topologically, MG-CPT cultivates spatial-relationship understanding, and TV-SFT preserves it during adaptation. Notably, even SPAR-4B outperforms all baselines on every metric, showing that spatial-perception enhancement yields greater gains than merely scaling parameters.

\subsubsection{Results on Industrial Datasets.}
Since GRMs are systematically superior, we compare SPAR with three SOTA GRMs. As shown in Table~\ref{exp:private}, SPAR maintains substantial advantages:



\textbf{SPAR maintains substantial advantages at industrial scale.} SPAR-8B outperforms PLUM on all four datasets, with a mean relative improvement of 37.30\% across the eight R@5/N@5 dataset--metric combinations; SPAR-4B and SPAR-0.6B achieve 27.38\% and 7.24\%, respectively. Notably, even the 0.6B variant surpasses every 8B baseline on all eight combinations. The comparison with PLUM is particularly informative because PLUM also adopts an SID--CPT--SFT pipeline~\cite{he2026plum}, whereas SPAR incorporates urban spatial knowledge into identifier construction, continued pre-training, and task adaptation. The consistent gains therefore support the benefits of spatially informed training beyond adopting a generic multi-stage pipeline, demonstrating its effectiveness across the evaluated industrial datasets.

\textbf{Model scaling continues to yield consistent gains.} Performance improves monotonically with backbone size, with the mean relative improvement over PLUM increasing from 7.24\% (0.6B) to 27.38\% (4B) and 37.30\% (8B). As shown in the final row of Table~\ref{exp:private}, scaling from 4B to 8B yields further relative improvements of 4.58\%--9.44\% across the eight dataset--metric combinations. These results show that the benefits of spatially informed training persist across model scales and that larger backbones further improve recommendation performance within SPAR.

\begin{table}[t]
\centering
\caption{Ablation study on industrial datasets. R.I(TV-SFT) is the relative improvement contributed by the TV-SFT module.}
\label{exp:ablation_performance}
\scalebox{0.70}{
    \begin{tabular}{cc|ccc|ccc|c} 
\hline\hline
 & Models    & R@5& R@10        & R@20       & N@5& N@10& N@20& Ave.Imp\\ 
\hline
\multirow{4}{*}{\rotatebox{270}{BeiJing}}
& TextOnlyGR& 0.4391& 0.5423& 0.6455& 0.3059& 0.3392& 0.3654& -\\
& SPAR-NoCPT& 0.4656& 0.5790& 0.6768& 0.3566& 0.3933& 0.4181& 10.77\%\\
& SPAR-FullSFT& \underline{0.5324}   & \underline{0.6429}   & \underline{0.7321}  & \underline{0.3975}   & \underline{0.4334}   & \underline{0.4560}   & \underline{22.62\%}   \\
& SPAR& \textbf{0.5708}  & \textbf{0.6813}  & \textbf{0.7702} & \textbf{0.4325}  & \textbf{0.4684}  & \textbf{0.4910}  & \textbf{31.47\%}  \\ 
\cline{2-9}
& \textbf{R.I(TV-SFT)} & \textbf{7.21\%}  & \textbf{5.97\%}  & \textbf{5.20\%} & \textbf{8.80\%}  & \textbf{8.07\%}  & \textbf{7.67\%}  & -\\ 
\hline

\multirow{4}{*}{\rotatebox{270}{ShangHai}}
& TextOnlyGR& 0.3706& 0.4794& 0.5861& 0.2693& 0.3045& 0.3315& -\\
& SPAR-NoCPT& 0.4056& 0.5165& 0.6164& 0.2907& 0.3267& 0.3520& 7.30\% \\
& SPAR-FullSFT& \underline{0.4376}   & \underline{0.5486}   & \underline{0.6537}  & \underline{0.3204}   & \underline{0.3564}   & \underline{0.3830}   & \underline{15.93\%}   \\
& SPAR& \textbf{0.4910}  & \textbf{0.5989}  & \textbf{0.6927} & \textbf{0.3656}  & \textbf{0.4007}  & \textbf{0.4245}  & \textbf{28.50\%}  \\ 
\cline{2-9}
& \textbf{R.I(TV-SFT)} & \textbf{12.20\%} & \textbf{9.17\%}  & \textbf{5.97\%} & \textbf{14.11\%} & \textbf{12.43\%} & \textbf{10.84\%} & -\\ 
\hline
\multirow{4}{*}{\rotatebox{270}{TianJin}}
& TextOnlyGR& 0.3801& 0.4881& 0.5961& 0.2646& 0.2996& 0.3269& -\\
& SPAR-NoCPT& 0.4050& 0.5149& 0.6184& 0.2948& 0.3304& 0.3567& 7.77\% \\
& SPAR-FullSFT& \underline{0.4380}   & \underline{0.5516}   & \underline{0.6571}  & \underline{0.3225}   & \underline{0.3593}   & \underline{0.3861}   & \underline{16.40\%}   \\
& SPAR& \textbf{0.5051}  & \textbf{0.6191}  & \textbf{0.7171} & \textbf{0.3775}  & \textbf{0.4145}  & \textbf{0.4394}  & \textbf{32.58\%}  \\ 
\cline{2-9}
& \textbf{R.I(TV-SFT)} & \textbf{15.32\%} & \textbf{12.24\%} & \textbf{9.13\%} & \textbf{17.05\%} & \textbf{15.36\%} & \textbf{13.80\%} & -\\ 
\hline

\multirow{4}{*}{\rotatebox{270}{ZheJiang}}
& TextOnlyGR& 0.4719& 0.5920& 0.6992& 0.3240& 0.3628& 0.3901& -\\
& SPAR-NoCPT& 0.4995& 0.6135& 0.7096& 0.3695& 0.4065& 0.4309& 7.92\%\\
& SPAR-FullSFT& \underline{0.5129}   & \underline{0.6269}   & \underline{0.7224}  & \underline{0.3797}   & \underline{0.4167}   & \underline{0.4409}   & \underline{10.50\%}   \\
& SPAR& \textbf{0.5424}  & \textbf{0.6506}  & \textbf{0.7379} & \textbf{0.4079}  & \textbf{0.4431}  & \textbf{0.4653}  & \textbf{16.28\%}  \\ 
\cline{2-9}
& \textbf{R.I(TV-SFT)} & \textbf{5.74\%}  & \textbf{3.77\%}  & \textbf{2.15\%} & \textbf{7.43\%}  & \textbf{6.33\%}  & \textbf{5.54\%}  & -\\
    \hline\hline
        \end{tabular}
    }
\end{table}

\subsection{Ablation Study}
To validate the progressive contributions of SPAR's core components, we conduct a systematic ablation study across four industrial datasets. As shown in Table~\ref{exp:ablation_performance}, we compare four variants that progressively incorporate additional components: TextOnlyGR (text-based SID + FullSFT), SPAR-NoCPT (SI-SID + FullSFT), SPAR-FullSFT (SI-SID + MG-CPT + FullSFT), and finally SPAR (SI-SID + MG-CPT + TV-SFT), observing the stepwise performance evolution. R.I(TV-SFT) denotes the relative improvement of SPAR over SPAR-FullSFT, isolating the contribution of TV-SFT. Based on Table~\ref{exp:ablation_performance}, we draw three key conclusions:

\textbf{SI-SID and MG-CPT provide complementary spatial benefits.} Replacing text-based identifiers with SI-SID increases the mean relative improvement over TextOnlyGR to 7.30\%--10.77\% across the four datasets, showing that geographically grounded identifiers benefit recommendation even without continued pre-training. Adding MG-CPT further increases this improvement by 2.58--11.85 percentage points. The gains from both stages support their complementary roles: SI-SID introduces explicit location information into identifier construction, while MG-CPT enables the backbone to learn spatial attributes and relationships from POI, road, and navigation data. This progression supports the design of establishing spatially grounded representations before learning urban spatial relations.

\textbf{TV-SFT further improves recommendation after MG-CPT.} Relative to SPAR-FullSFT, TV-SFT yields improvements of 2.15\%--17.05\% across the four datasets, with Tianjin showing the largest gains at R@5 and N@5 (15.32\% and 17.05\%, respectively). Although MG-CPT followed by full-parameter SFT already improves performance, task-vector anchored adaptation provides further gains on every reported metric. These results are consistent with the motivation of retaining a fixed CPT-induced parameter component while allowing behavioral adaptation through the trainable base branch and low-rank correction. The comparison supports TV-SFT as an effective adaptation strategy for spatially informed POI recommendation.

\textbf{Cumulative advantage of the full SPAR framework.} The complete framework improves over TextOnlyGR by 16.28\%--32.58\% across the four datasets, and each successive stage improves all reported metrics. These consistent gains support the complementary contributions of SI-SID, MG-CPT, and TV-SFT: spatially grounded identifiers provide the representation foundation, MG-CPT adds urban spatial learning, and TV-SFT improves downstream adaptation. Together, the results demonstrate the effectiveness of incorporating spatial knowledge into both representation construction and model training for industrial POI recommendation.

\subsection{Spatial Proximity and Evolution Analysis}

\begin{figure}[t]
    \centering
        \subfloat[Shanghai]{
\begin{minipage}[t]{0.32\linewidth}
\centering
\includegraphics[width=0.95\textwidth]{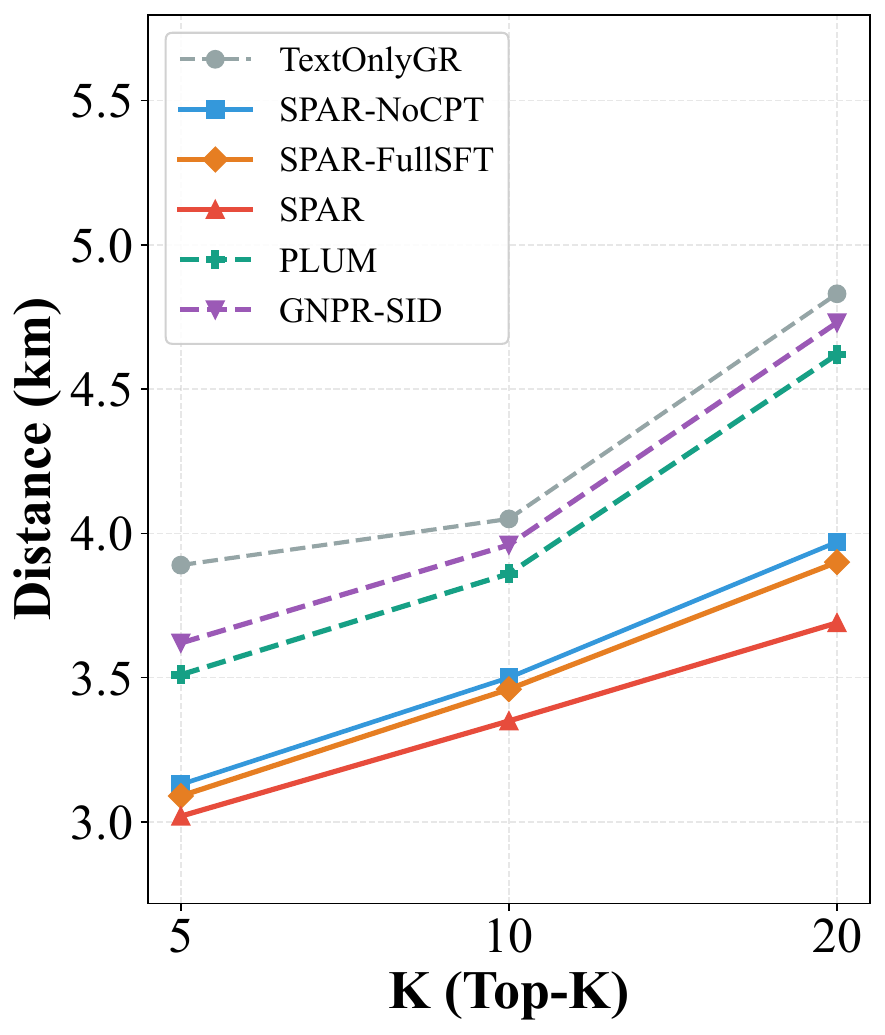}
\end{minipage}
        }
        \subfloat[Tianjin]{
\begin{minipage}[t]{0.32\linewidth}
\centering
\includegraphics[width=0.95\textwidth]{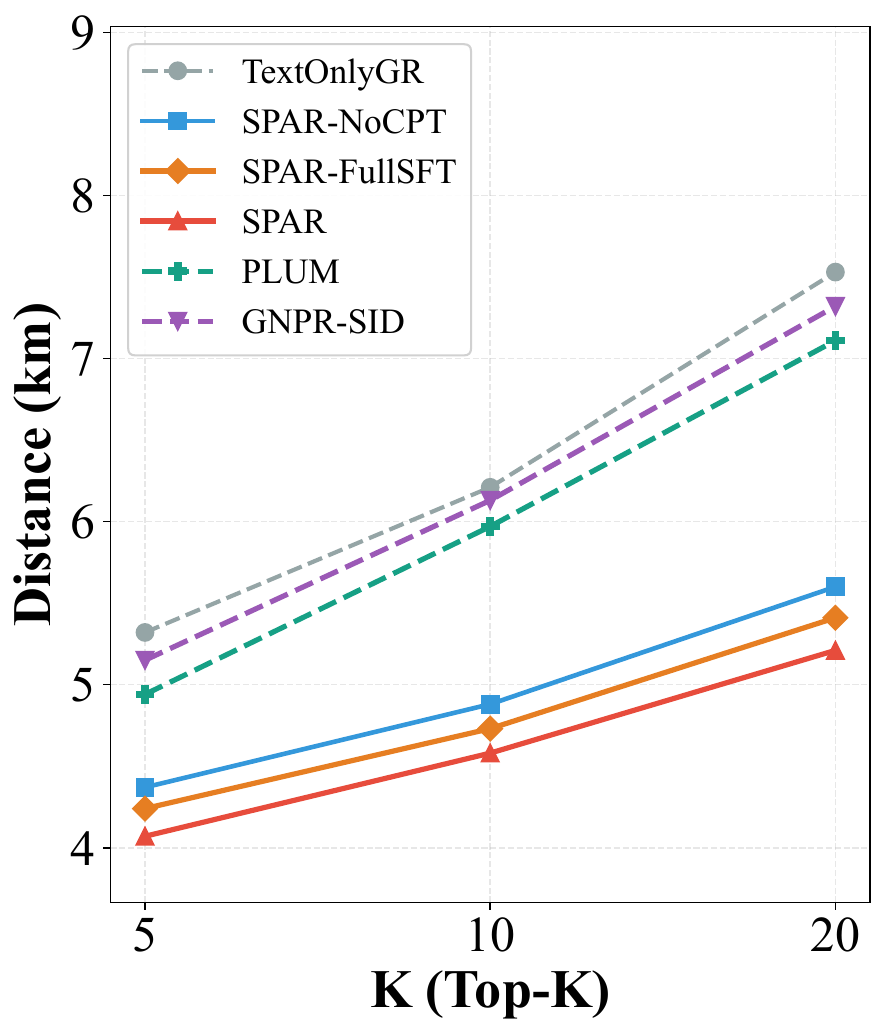}
\end{minipage}
        }
        \subfloat[Incremental Imp.]{
\begin{minipage}[t]{0.32\linewidth}
\centering
\includegraphics[width=0.95\textwidth]{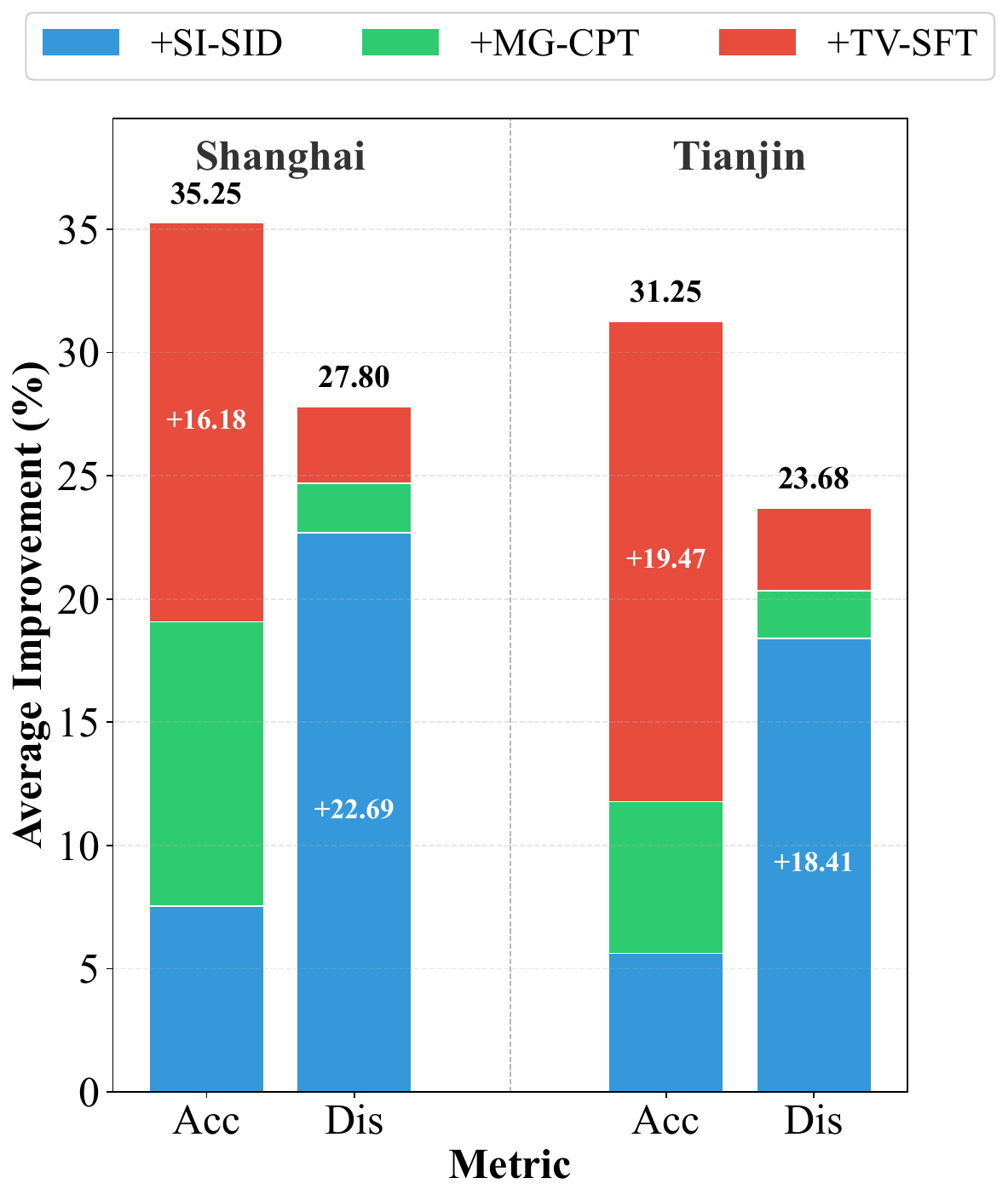}
\end{minipage}
        }
        \caption{(a)/(b): distance comparison across Top-K recommended POIs to user's location; (c): Incremental contribution to distance and accuracy average improvement.}
        \label{exp:abl_distance}
\end{figure}

We examine how SI-SID affects spatial proximity and how each stage contributes to SPAR. Figure~\ref{exp:abl_distance}(a)/(b) plot the average haversine distance between the Top-K recommended POIs and the user's real-time location (represented by Dis@K), while Figure~\ref{exp:abl_distance}(c) decomposes the incremental relative improvement (Ave.Imp over TextOnlyGR, from Table~\ref{exp:ablation_performance}) contributed by sequentially adding SI-SID, MG-CPT, and TV-SFT. We draw three key conclusions:

\textbf{(1) SI-SID establishes the spatial foundation by pulling recommended POIs closer to the user. }TextOnlyGR, GNPR-SID, and PLUM all derive identifiers from textual descriptions containing coordinates and addresses, further reinforced by contrastive objectives and collaborative signals, yet their Dis@5 remains 3.51--3.89 km on Shanghai and 4.94--5.32 km on Tianjin. SI-SID instead fuses normalized geospatial features into the representation before quantization, making proximity a structural property of the codebook. SPAR-NoCPT, which differs from these baselines mainly in adopting SI-SID, already lowers Dis@5 to 3.13 and 4.37 km, with accuracy gains of 7.30\% and 7.77\%. SI-SID thus provides the spatial foundation on which MG-CPT and TV-SFT further build.

\textbf{(2) MG-CPT and TV-SFT are the main drivers of relative improvement.} Adding MG-CPT (SPAR-NoCPT $\rightarrow$ SPAR-FullSFT) lifts relative improvement from 7.30\% to 15.93\% on Shanghai and 7.77\% to 16.40\% on Tianjin, and the final TV-SFT stage brings the largest gains (28.50\% and 32.58\%), while distance keeps decreasing (to 27.80\% and 23.68\% cumulative reduction). Explicit spatial-reasoning cultivation and knowledge-preserving adaptation are thus the core mechanisms that turn spatial awareness into recommendation quality.

\textbf{(3) The stepwise evolution validates the SPAR design.} Distance reduction is led by SI-SID while accuracy gains are led by MG-CPT and TV-SFT: SI-SID embeds spatial topology into the identifier vocabulary, upon which MG-CPT and TV-SFT cultivate and preserve spatial reasoning. This division is consistent with our observation: coordinate-infused SI-SID is the foundation and consistently reinforcing real urban spatial knowledge is the core driver of effective generative POI recommendation.

\subsection{Cold-Start Performance}
To evaluate SPAR's robustness under data-sparse conditions, we split each dataset by user behavior sequence length into cold-start ($<8$) and warm-start ($\ge 16$) subsets; note that query terms and real-time location remain available in both, and only historical interactions are sparse. Table~\ref{exp:cold_start_result} summarizes the performance.


\textbf{Sparse behavioral histories are associated with lower performance.} TextOnlyGR suffers substantial performance degradation when behavior histories are sparse, with Recall@5 falling 27.8\% on Shanghai (0.3908 → 0.2821) and 38.4\% on Zhejiang (0.4922 → 0.3033), with similar declines in NDCG@5, confirming that behavior sequences are a critical signal for POI recommendation.

\textbf{SPAR provides larger gains under sparse behavioral histories.} SPAR improves over TextOnlyGR in both settings, with larger relative gains in cold-start Recall@5: 47.6\% on Shanghai and 48.0\% on Zhejiang, compared with 30.7\% and 14.9\% in warm-start. These larger gains suggest that spatially informed representations and training provide useful complementary signals when behavioral histories are limited. SPAR also narrows the relative cold-to-warm Recall@5 gap from 27.8\%--38.4\% for TextOnlyGR to 18.4\%--20.6\%, supporting its robustness under sparse-history conditions.

\begin{table}
    \centering
    \caption{Summary of cold- and warm-start performance.}
    \label{exp:cold_start_result}
    \setlength{\tabcolsep}{2.5mm}
    \scalebox{0.80}{
    \begin{tabular}{cc|cc|cc}
    \hline
    \hline
    \multicolumn{2}{c|}{Datasets} & \multicolumn{2}{c|}{Shanghai} & \multicolumn{2}{c}{Zhejiang} \\
    \hline
    Task & Metrics & TextOnlyGR & SPAR & TextOnlyGR & SPAR \\
    \hline
    \multirow{2}{*}{Cold-start} & R@5 & 0.2821 & \textbf{0.4164} & 0.3033 & \textbf{0.4490} \\
     & N@5 & 0.2107 & \textbf{0.3157} & 0.2442 & \textbf{0.3286} \\
    \hline
    \multirow{2}{*}{Warm-start} & R@5 & 0.3908 & \textbf{0.5106} & 0.4922 & \textbf{0.5654} \\
     & N@5 & 0.2950 & \textbf{0.3773} & 0.3512 & \textbf{0.4217} \\
    \hline
    \hline
    \end{tabular}
    }
\end{table}

\subsection{Hierarchical Visualization of SI-SID} 
\begin{figure}[t]
    \setlength{\abovecaptionskip}{0.2cm}
    \setlength{\belowcaptionskip}{-0.4cm}
    \centering
        \subfloat[Semantic embedding t-SNE.]{
\begin{minipage}[t]{0.49\linewidth}
\centering
\includegraphics[width=0.90\textwidth]{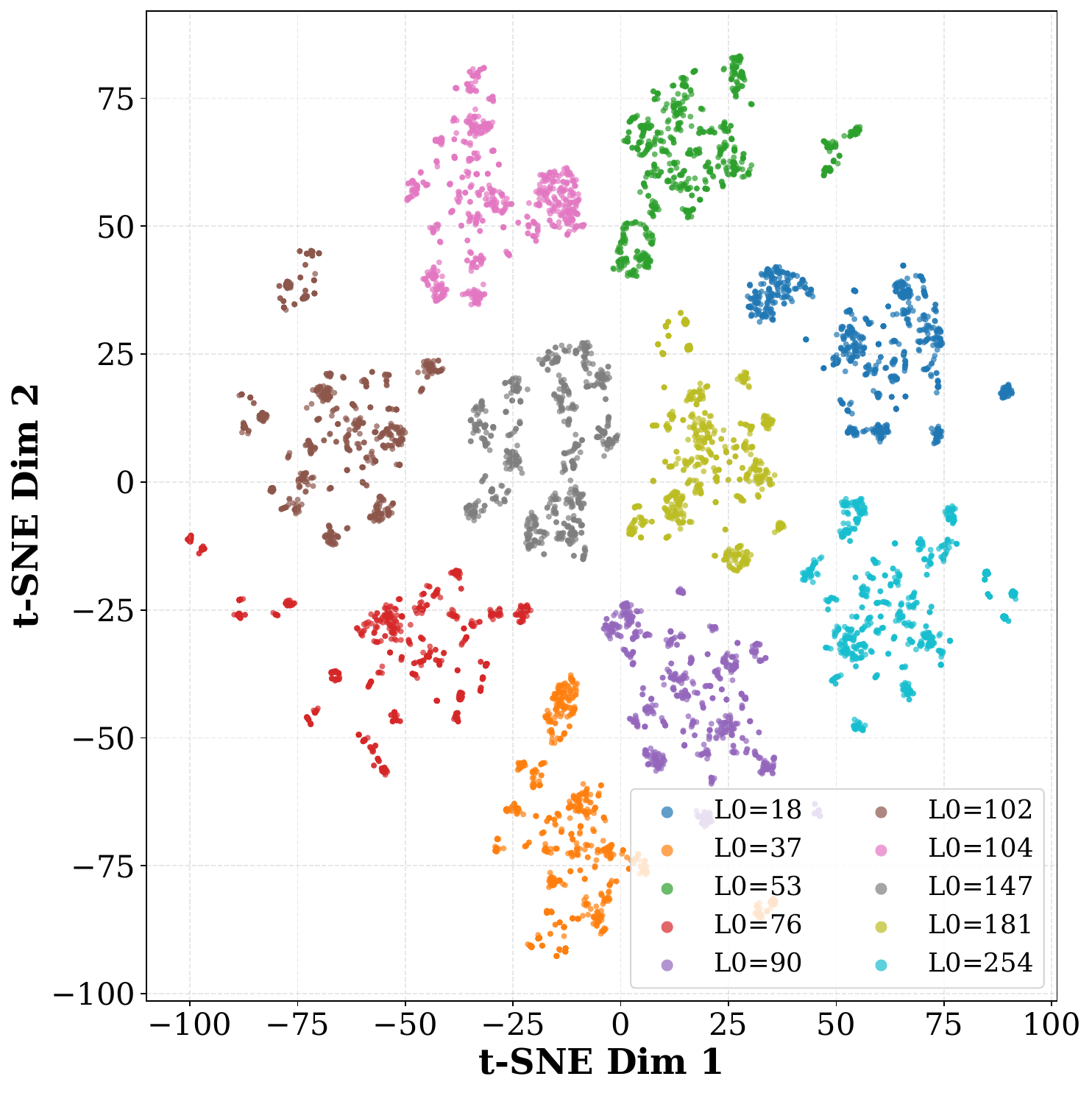}
\end{minipage}
        }
        \subfloat[Geo coordinates.]{
\begin{minipage}[t]{0.49\linewidth}
\centering
\includegraphics[width=0.99\textwidth]{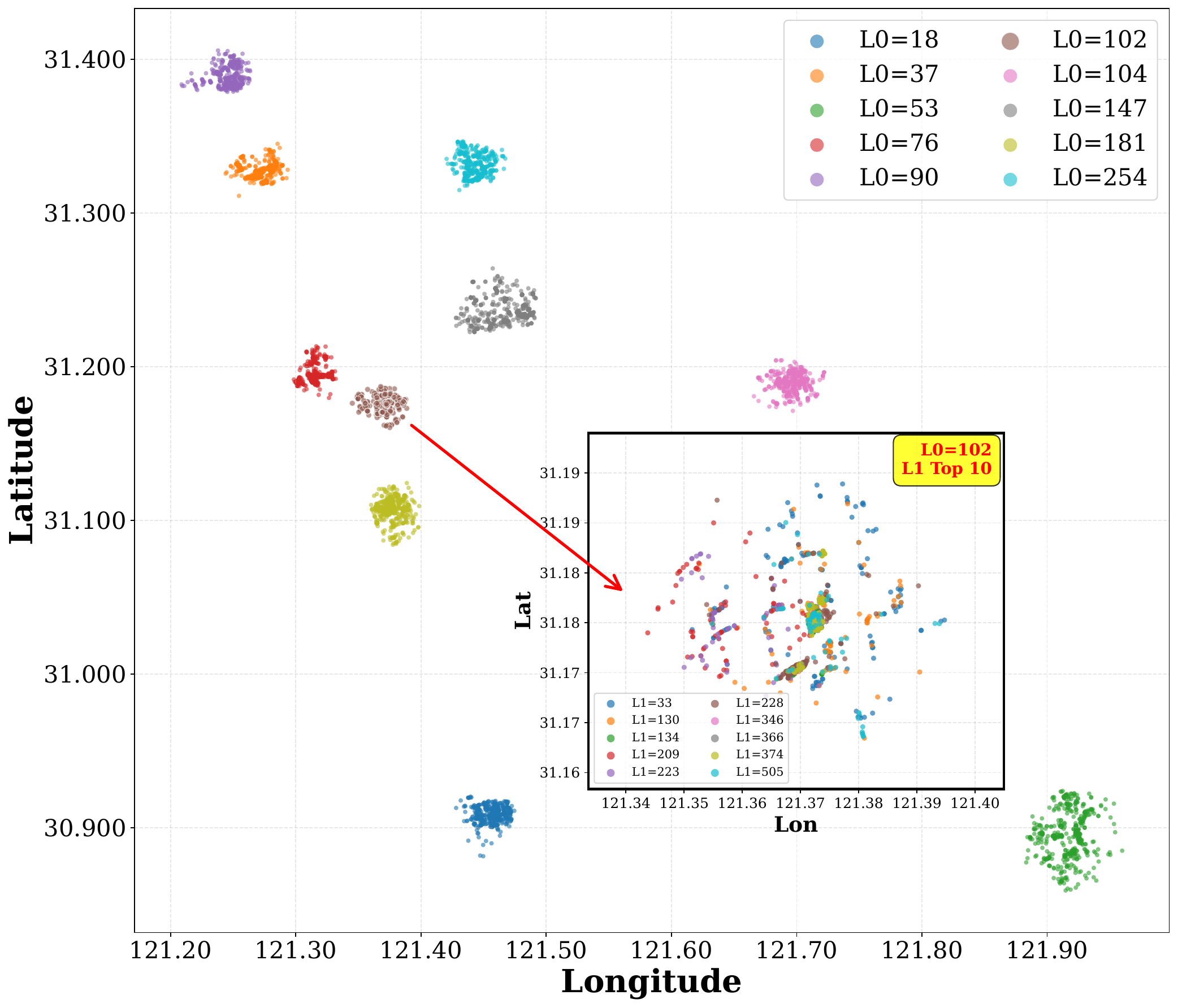}
\end{minipage}
        }
        
        \subfloat[Semantic similarity heatmap]{
\begin{minipage}[t]{0.49\linewidth}
\centering
\includegraphics[width=0.95\textwidth]{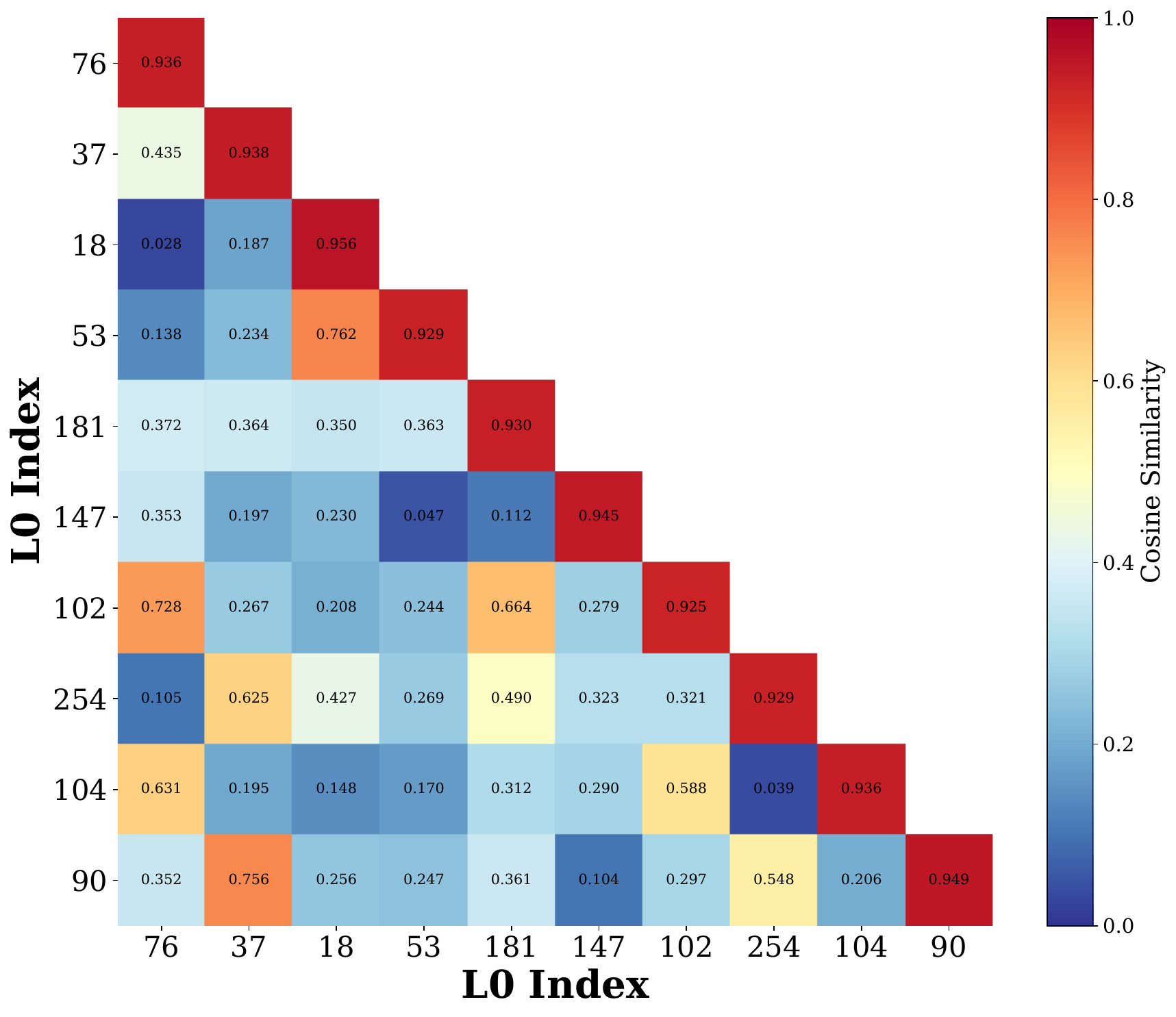}
\end{minipage}
        }
        \subfloat[Semantic t-SNE in L0=102 cluster]{
\begin{minipage}[t]{0.49\linewidth}
\centering
\includegraphics[width=0.90\textwidth]{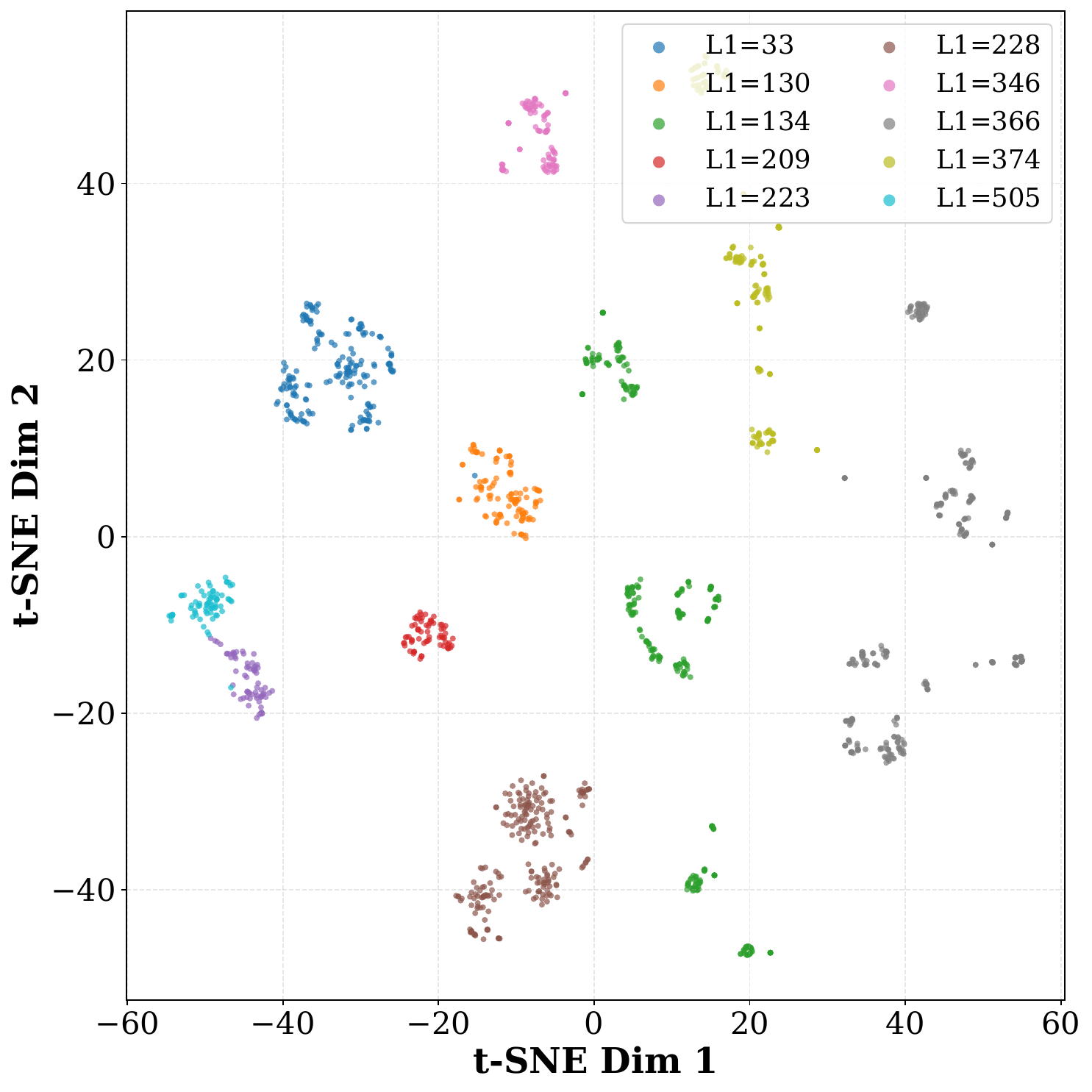}
\end{minipage}
        }
        \caption{Multi-view analysis of SI-SID clusters reveals hierarchical semantic and geographic structure.}
        \label{fig:sisid_exps}
\end{figure}

To understand how SI-SID organizes POIs in a hierarchical semantic-geographic space, we conduct a multi-view analysis of its quantization structure: Figure \ref{fig:sisid_exps}(a) and (b) show the semantic t-SNE and geographic coordinates of the top-10 L0 clusters, Figure \ref{fig:sisid_exps}(c) quantifies their intra- and inter-cluster semantic similarity, and Figure \ref{fig:sisid_exps}(d) zooms into L0=102 to visualize its top-10 L1 sub-clusters (geographic distribution in the inset of Figure \ref{fig:sisid_exps}(b)). We highlight two important conclusions:

\textbf{L0 clusters exhibit both semantic consistency and geographic distinguishability.} Figure \ref{fig:sisid_exps}(a) shows that POIs within the same L0 cluster form compact groups in the semantic t-SNE space, while Figure \ref{fig:sisid_exps}(c) confirms high intra-cluster semantic similarity. Simultaneously, Figure \ref{fig:sisid_exps}(b) reveals that different L0 clusters occupy distinct geographic regions with minimal spatial overlap. This indicates that the first-level quantization primarily partitions POIs based on geographic proximity: nearby venues tend to fall into the same L0 cluster, and their semantic consistency likely arises because adjacent urban regions often share similar business compositions or functional roles.

\textbf{L1 sub-clusters capture fine-grained semantic distinctions beyond geographic proximity.} Within L0=102, the top-10 L1 sub-clusters separate into distinct semantic groups (Figure \ref{fig:sisid_exps}(d)) yet overlap geographically (inset of Figure \ref{fig:sisid_exps}(b)). Since POIs in an L0 cluster are already geographically proximate, the geographic dimension is largely exhausted at the first level; the residual passed to L1 is therefore dominated by semantic variation, and L1 naturally refines these POIs by semantic category rather than location. This coarse-to-fine progression, geographic partitioning at L0 followed by semantic refinement at L1, confirms the effectiveness of SI-SID tokenization, allowing SPAR to combine the coarse geographic constraint from L0 with fine-grained semantic preferences for retrieval conditioned on user location and query intent.

\subsection{Spatial Cognition Evaluation for MG-CPT}
\label{sec:mgcpt_exp}

\begin{figure}[t]
    \centering
        \subfloat[Spatial cognition evaluation on Shanghai Dataset]{
\begin{minipage}[t]{0.99\linewidth}
\centering
\includegraphics[width=0.99\textwidth]{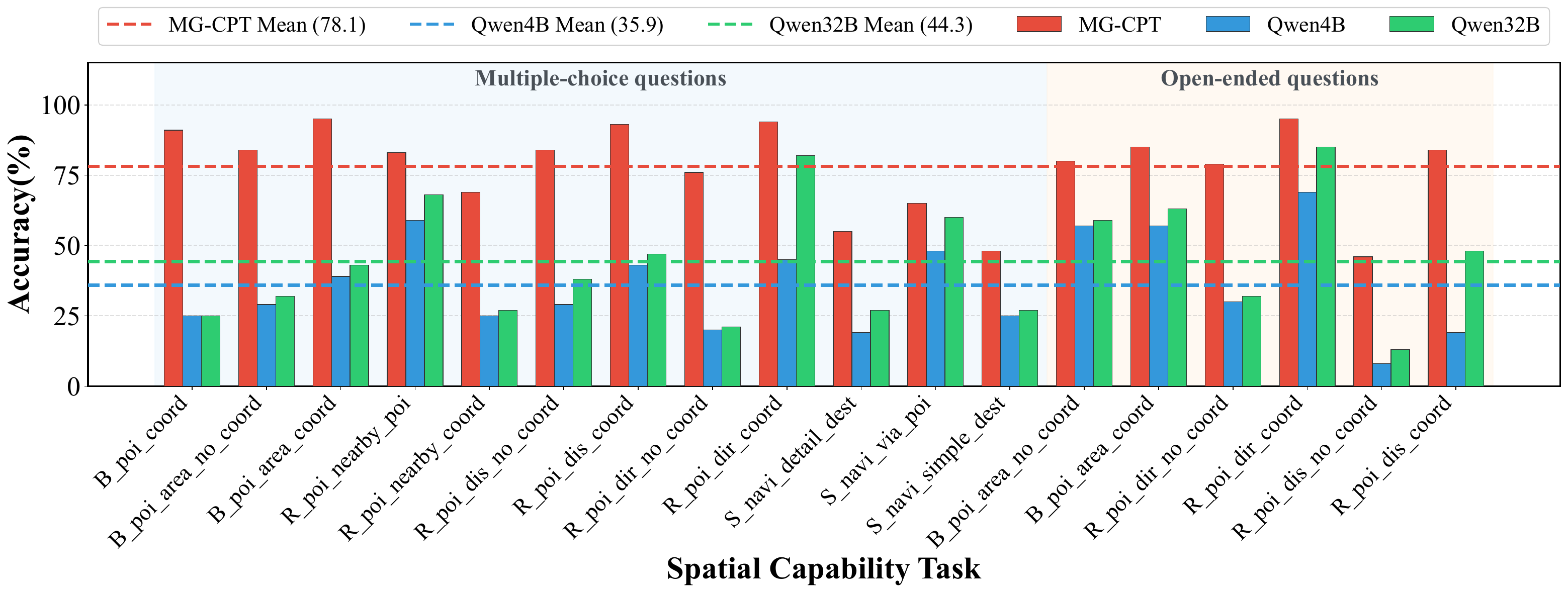}
\end{minipage}
        }
        
        \subfloat[Spatial cognition evaluation on Beijing Dataset]{
\begin{minipage}[t]{0.99\linewidth}
\centering
\includegraphics[width=0.99\textwidth]{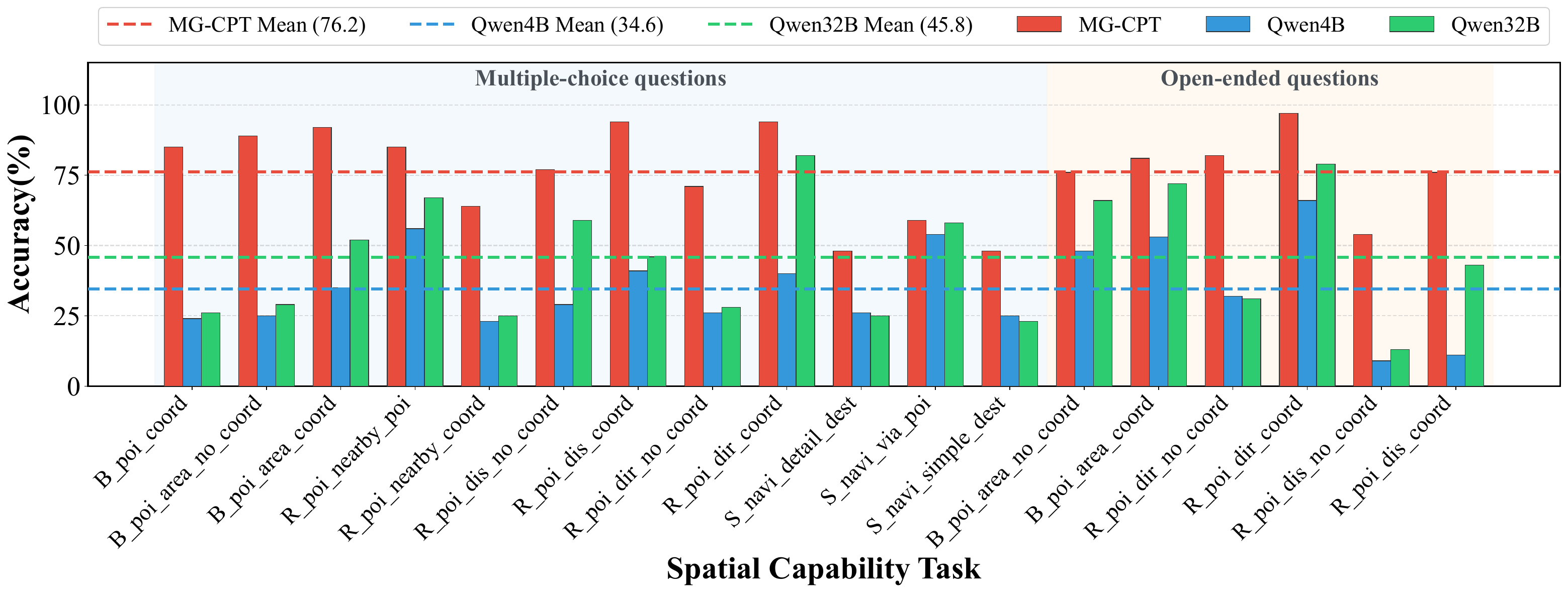}
\end{minipage}
        }
    \caption{Spatial cognition evaluation across 18 tasks.}
\label{exp:mgcpt}
\end{figure}

To assess whether MG-CPT effectively acquires spatial cognition, we construct one benchmark for each industrial dataset, each consisting of 18 tasks with 2{,}000 questions, spanning multiple-choice and open-ended formats; Figure~\ref{exp:mgcpt} reports results on Shanghai and Beijing. Most numbers in the subsection are accuracies in percent, averaged over the 18 spatial tasks of each benchmark. Each task name combines a category prefix---\texttt{B}, \texttt{R}, \texttt{S} for basic-attribute, relational, and systemic questions---with an ability suffix: \texttt{dis}/\texttt{dir} for distance/direction reasoning, \texttt{area}/\texttt{nearby} for region/surroundings queries, and \texttt{navi} for navigation. The \texttt{coord}/\texttt{no\_coord} suffix indicates whether coordinates are provided or omitted, with the latter forcing reliance on internalized geographic knowledge. Based on these results, we draw several key conclusions:

\textbf{Targeted MG-CPT yields greater gains than parameter scaling alone.} On the Shanghai benchmark, MG-CPT increases the mean task accuracy of Qwen3-4B from 35.9\% to 78.1\%, a 42.2-percentage-point improvement that more than doubles the base accuracy. It also surpasses the eight-times-larger Qwen3-32B (44.3\%) by 33.8 percentage points. This comparison shows that targeted training on the 25 structured geospatial datasets provides substantial benefits beyond increasing parameter count on the evaluated city-specific tasks.

\textbf{General-model scaling provides smaller gains on city-specific spatial tasks.} Increasing the backbone from 4B to 32B raises the mean accuracy on Shanghai from 35.9\% to 44.3\%, an 8.4-percentage-point improvement. By comparison, the gain from MG-CPT on the 4B backbone is about five times as large. These results suggest that larger general-purpose models alone remain limited on the evaluated city-specific tasks, highlighting the value of explicit geospatial training.



\textbf{MG-CPT supports spatial questions without explicit coordinates.} Base LLMs depend heavily on explicit coordinates: Qwen3-32B, for example, jumps from 21 to 82 on directional questions once coordinates are provided. In contrast, MG-CPT remains strong even without coordinates (76 and 79 on directional Choice and Open questions), exceeding the coordinate-free Qwen3-32B by more than 3$\times$. This indicates that MG-CPT encodes the geospatial configuration of the corresponding city into its parameters, enabling metric and directional reasoning without coordinates supplied at inference time.

\textbf{Performance follows the three-tier complexity gradient.} Accuracy is highest on Basic-attribute tasks (80--95), moderate on Relational tasks, and lowest on Systemic navigation tasks (48--65), mirroring the increasing spatial complexity by which the training data are organized. Although navigation remains the hardest tier, MG-CPT still substantially outperforms both baselines on it, showing that even city-scale connectivity is partially internalized.

\subsection{Case Study: Proximity versus Reachability}
\begin{figure}[t]
    \setlength{\abovecaptionskip}{0.2cm}
    \setlength{\belowcaptionskip}{-0.4cm}
    \centering
\centering
\includegraphics[width=0.44\textwidth]{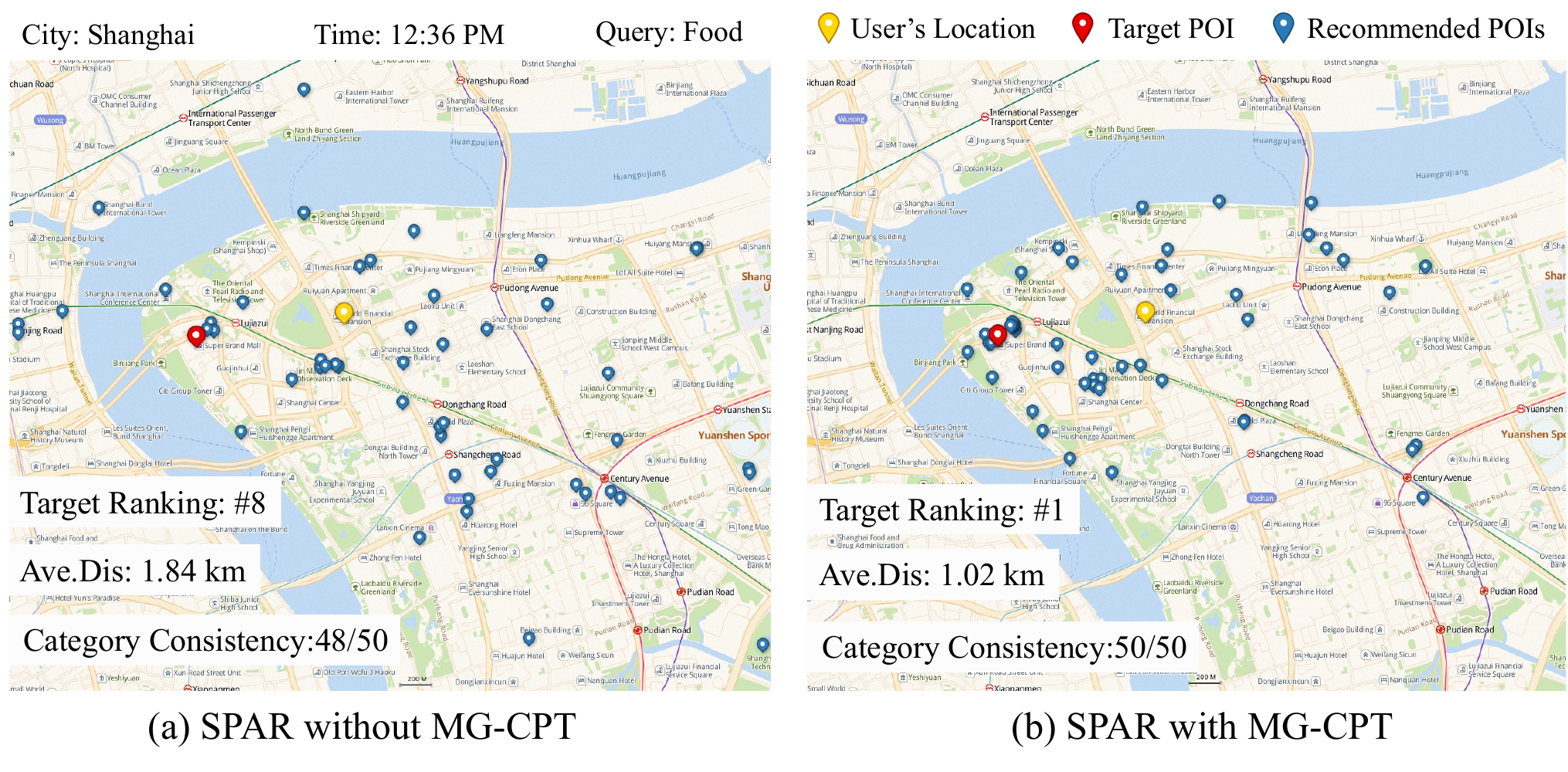}
    \caption{Visualization of recommended POIs distribution.}
\label{fig:case_study}
\end{figure}

Figure~\ref{fig:case_study} visualizes the top-50 POIs recommended by SPAR \textbf{with and without MG-CPT} for a user located in Shanghai, together with the average distance to the user's location, the category consistency with the target, and the target's ranking. 

As shown in Figure~\ref{fig:case_study}(a), without MG-CPT, the recommendations already have high category consistency, with 48 of the 50 POIs matching the target category, but are spatially scattered: the average distance is 1.84 km and the target is ranked \#8. Several candidates lie on the opposite bank of the river, illustrating how straight-line proximity can differ from access through the road network.


In contrast, with MG-CPT (Figure~\ref{fig:case_study}(b)), the recommendations concentrate on the user's own side of the river along nearby roads, reducing the average distance to 1.02 km (a 44.6\% reduction), increasing category consistency to 50 of 50, and promoting the target to \#1. Since category consistency is already high in both variants, this example suggests that MG-CPT improves the spatial coherence of recommendations rather than merely changing their category composition. The same-bank concentration is consistent with the goal of accounting for road-network constraints beyond straight-line proximity, complementing the distance and ablation analyses above.


\subsection{Hyper-Parameters Analysis of TV-SFT} 
\begin{figure}[t]
    \setlength{\abovecaptionskip}{0.2cm}
    \setlength{\belowcaptionskip}{-0.4cm}
    \centering
        \subfloat[Rank r]{
\begin{minipage}[t]{0.45\linewidth}
\centering
\includegraphics[width=0.99\textwidth]{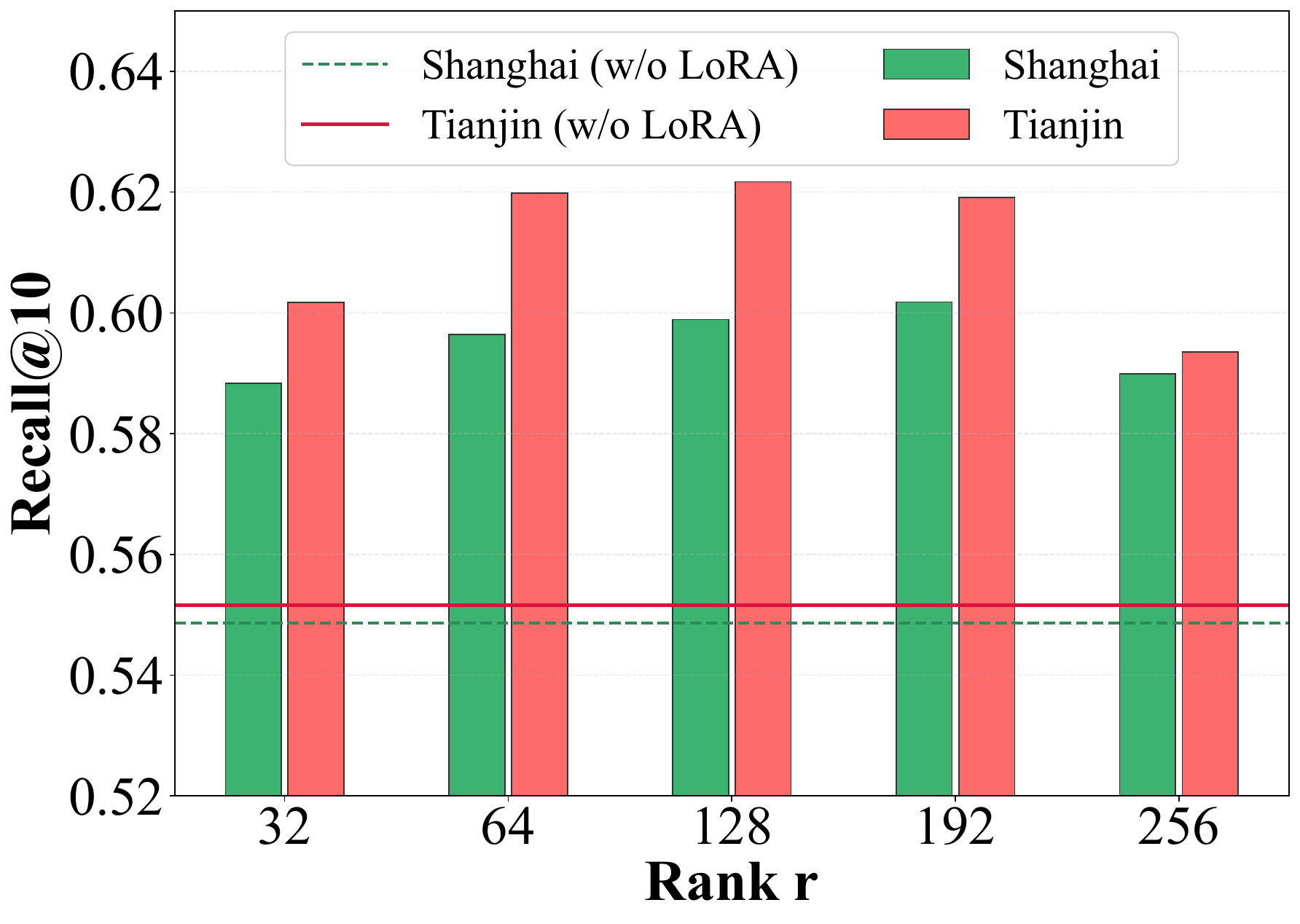}
\end{minipage}
        }
        \subfloat[Weight $\alpha$]{
\begin{minipage}[t]{0.45\linewidth}
\centering
\includegraphics[width=0.99\textwidth]{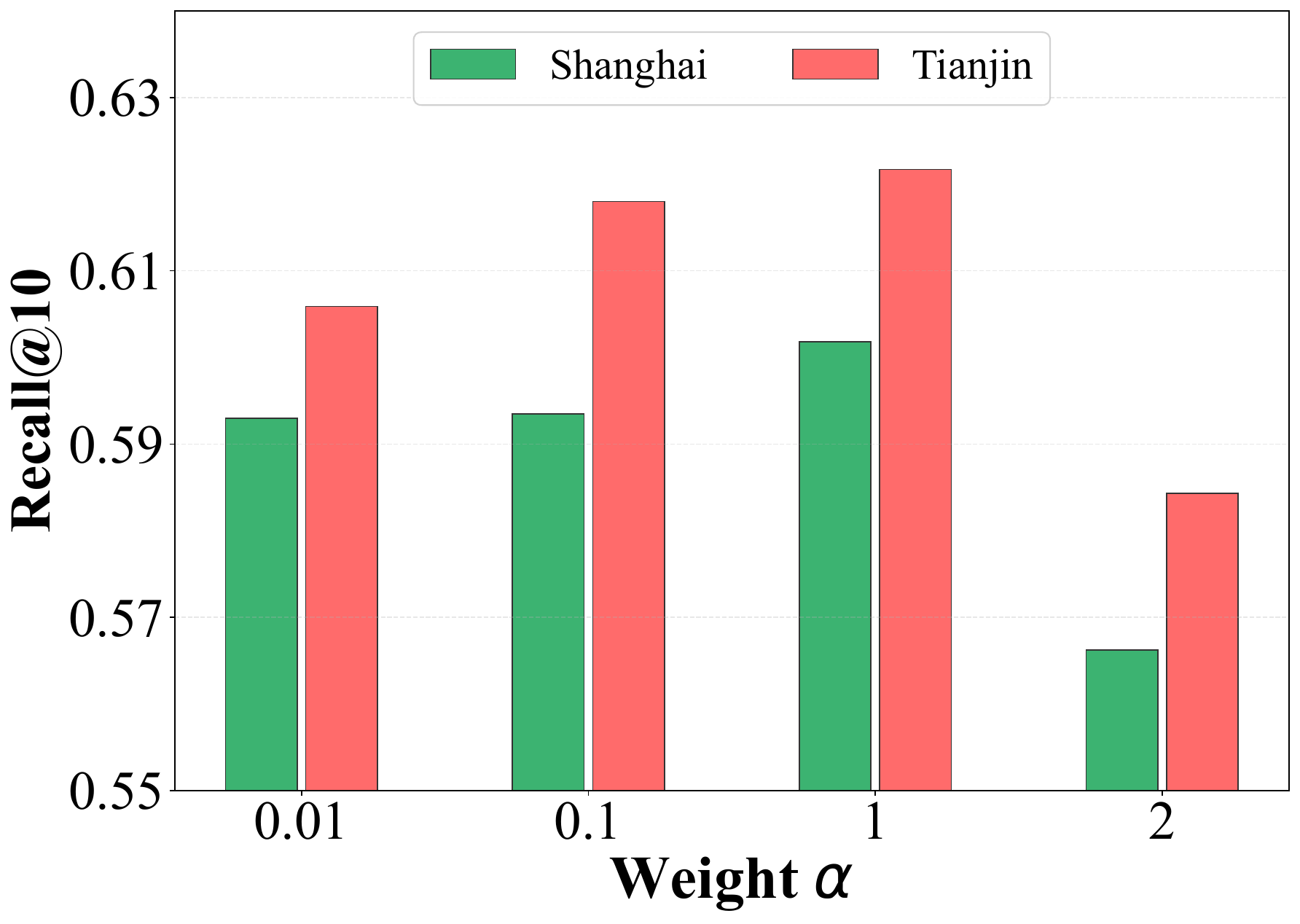}
\end{minipage}
        }
    \caption{Analysis of two hyper-parameters in TV-SFT.}
\label{exp:fig_hp}
\end{figure}

As shown in Figure~\ref{exp:fig_hp}, with all other settings fixed, we investigate two key hyperparameters of the TV-SFT stage: LoRA rank $r \in \{32, 64, 128, 192, 256\}$ and fusion weight $\alpha \in \{0.01, 0.1, 1, 2\}$. 

\textbf{LoRA Rank $r$.} The LoRA adapter refines the spatial knowledge carried by $\tau_{MG-CPT}$ so that it better fits the full-parameter branch of TV-SFT. Adding LoRA consistently outperforms the variant without it (horizontal reference line in Figure~\ref{exp:fig_hp}(a)), and although the optimal rank differs slightly across datasets (r=192 in Shanghai, r = 128 in Tianjin), performance varies within a narrow band over all tested ranks, indicating that the low-rank constraint confines spatial refinement to a robust subspace.



\textbf{Fusion Weight $\alpha$.} The fusion weight $\alpha$ controls the contribution of the frozen CPT task vector together with the trainable low-rank correction. As shown in Figure~\ref{exp:fig_hp}(b), both datasets achieve their best performance at $\alpha=1$, with lower performance at the other tested values. This result supports the default setting of $\alpha=1$, which retains the original scale of the CPT-induced parameter update while allowing adaptation through the trainable components.



\section{Conclusion}

In this paper, we argue that generative POI recommendation in LBS benefits from grounding behavioral preference modeling in real urban spatial relations. We propose SPAR, which incorporates urban spatial knowledge through three complementary stages. SI-SID constructs spatially grounded identifiers by combining coordinate encodings with semantic embeddings; MG-CPT cultivates spatial knowledge through continued pre-training on 25 geospatial datasets derived from POI, road, and navigation data; and TV-SFT retains the CPT-induced parameter component as a fixed anchor while adapting to user behavior. Experiments on two public benchmarks and four industrial datasets show that SPAR outperforms the evaluated baselines, with ablations supporting the contribution of each stage. Spatial cognition evaluations, sparse-history comparisons, and qualitative analyses further demonstrate improved city-specific spatial performance, robustness under limited behavioral histories, and spatially coherent recommendations. These results support urban spatial knowledge as a useful complement to behavioral modeling in generative POI recommendation.


\bibliographystyle{ACM-Reference-Format}
\bibliography{gr_bib}






\appendix

\begin{center}
  \LARGE\bfseries Appendices
\end{center}
\section{Theoretical Analysis of the Geospatial Encoder}
\label{app:geoencoder}

This appendix complements Section 4.1 Spatially-Intrinsic SID by analyzing why the sinusoidal geospatial encoder produces embeddings whose geometry faithfully mirrors geographic topology, the property that SI-SID inherits through residual quantization. We first state two exact properties of the Fourier features $\gamma(\cdot)$, then empirically verify that both properties survive the learned projection.

\begin{figure*}[t]
    \centering
        \subfloat[Original coordinate grid.]{
         \begin{minipage}[t]{0.38\linewidth}
         \centering
         \includegraphics[width=0.90\textwidth]{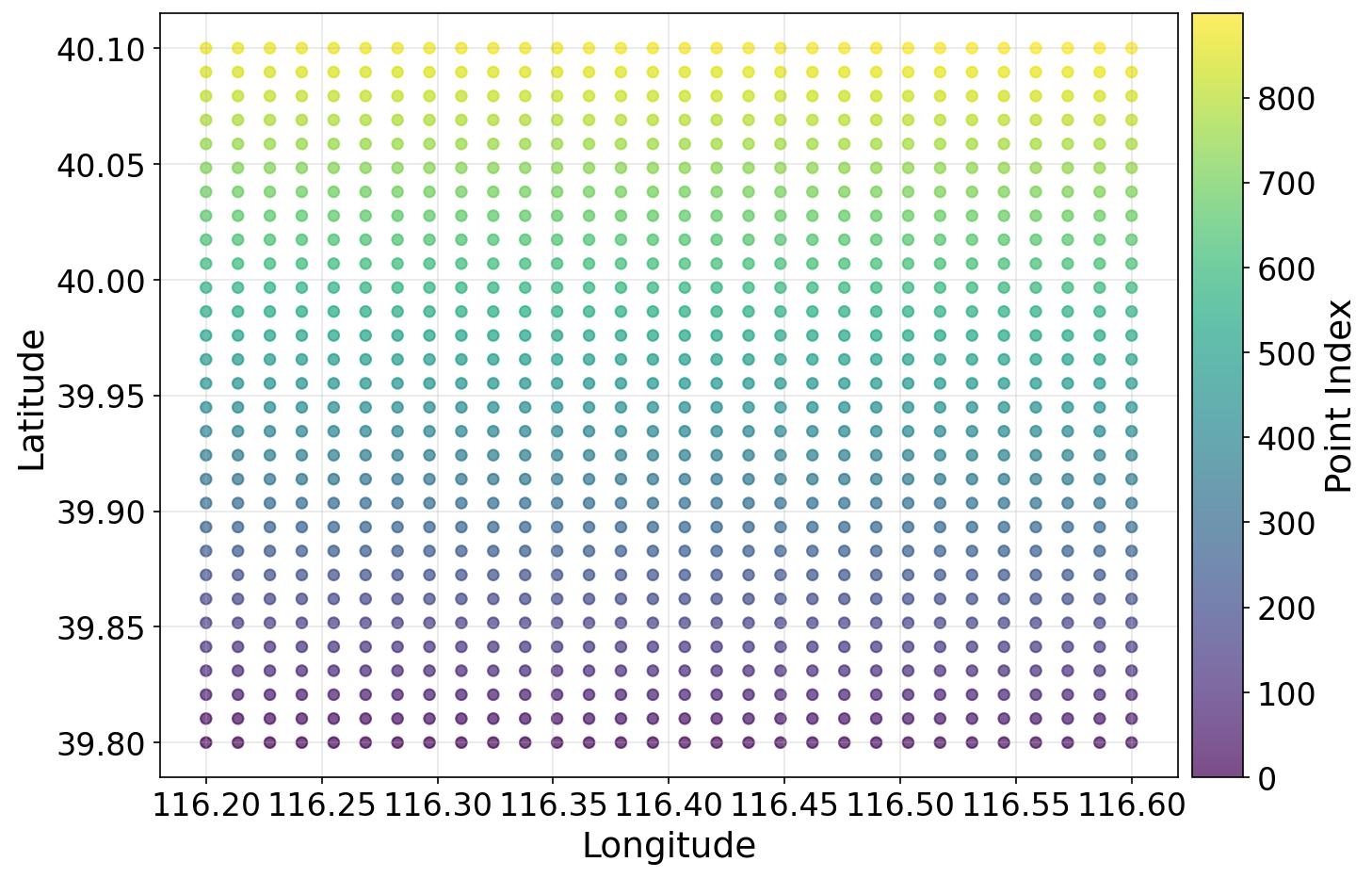}
         \end{minipage}
        }
        \subfloat[t-SNE of the geospatial embeddings.]{
         \begin{minipage}[t]{0.29\linewidth}
         \centering
         \includegraphics[width=0.99\textwidth]{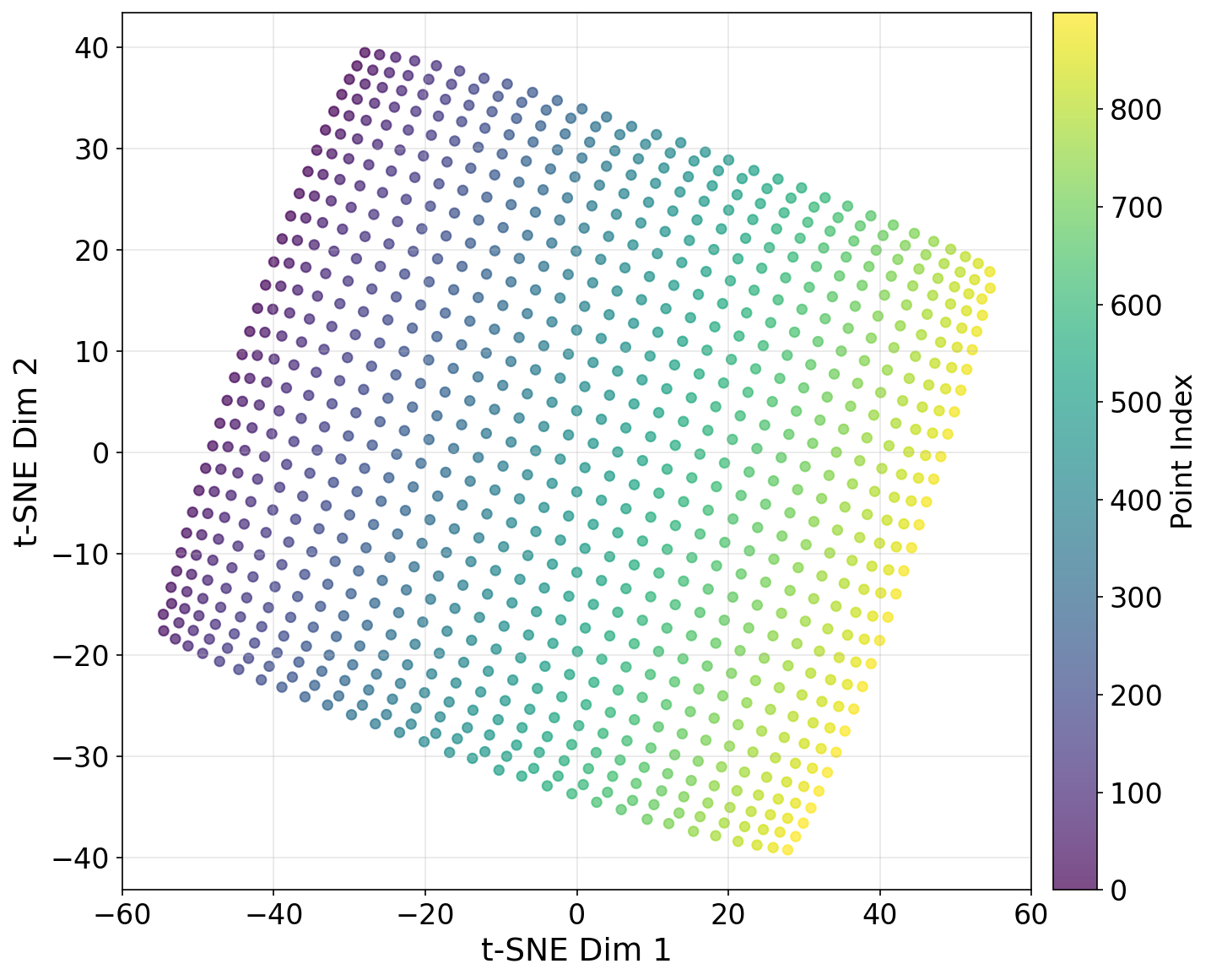}
         \end{minipage}
        }
        \subfloat[Embedding vs. geographic distance]{
         \begin{minipage}[t]{0.3\linewidth}
         \centering
         \includegraphics[width=0.95\textwidth]{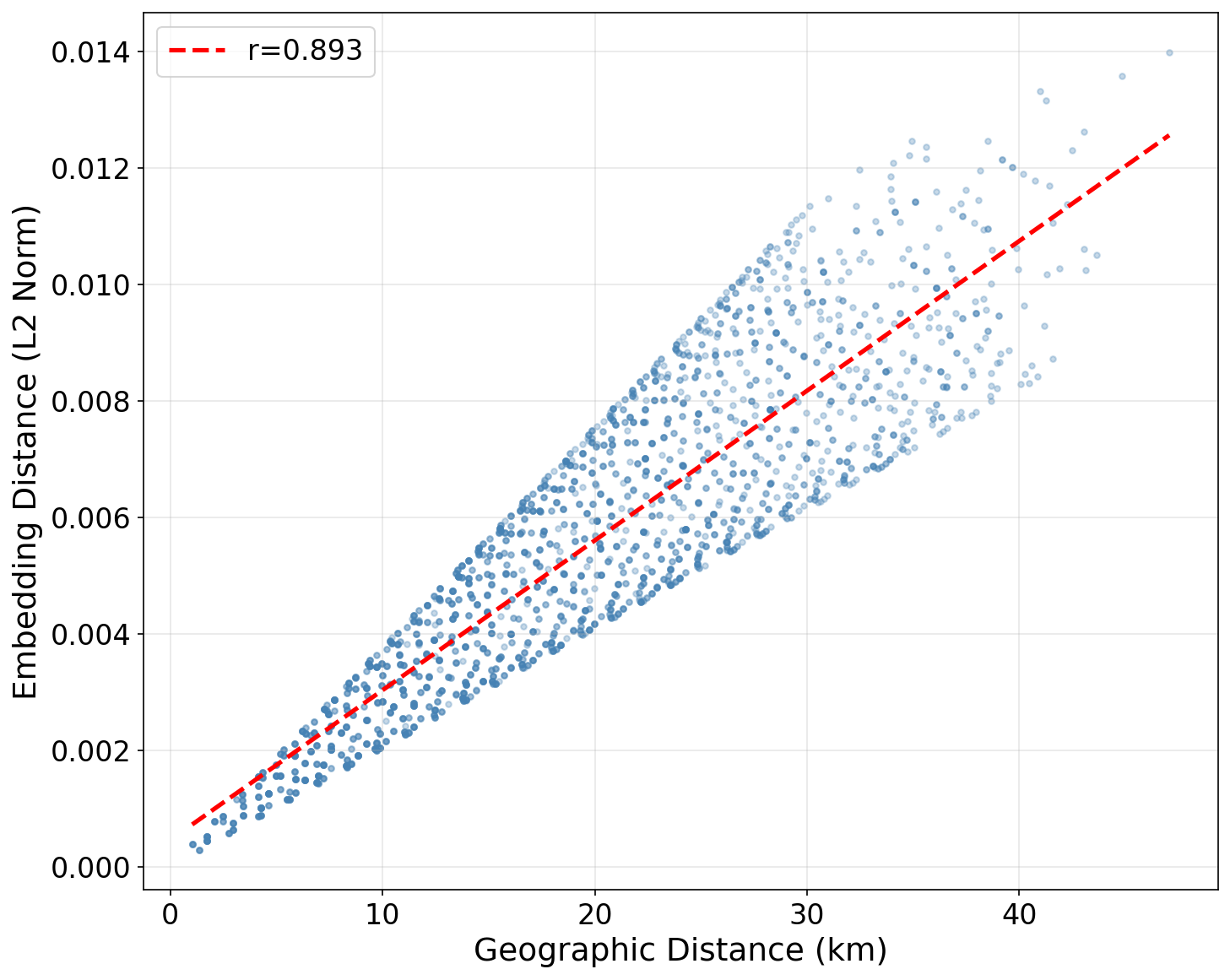}
         \end{minipage}
        }
        \caption{Empirical verification of the geospatial encoder on a $30\times30$ coordinate grid over Beijing. (a) The original grid, color-coded by point index. (b) The t-SNE projection of the embeddings preserves the grid topology: geographically adjacent points remain adjacent. (c) The embedding distance grows monotonically with the geographic distance (Pearson $r = 0.893$), confirming that the city's metric structure is faithfully carried into the embedding space.}
        \label{fig:geo_property}
\end{figure*}


\subsection{Exact Properties}

In what follows, let $u \in [0, 2\pi]$ denote a generic normalized coordinate (either $\hat{\lambda}$ or $\hat{\phi}$ from Section 4.1); the two properties hold for each coordinate separately and extend to their concatenation.

\textbf{Property 1 (Shift-invariant kernel).} For any two values $u_1, u_2$, the product-to-sum identity gives
\begin{equation}
    \begin{aligned}
     \langle \gamma(u_1), \gamma(u_2) \rangle 
     &= \sum_{k=0}^{K-1} \big[ \sin(f_k u_1)\sin(f_k u_2) + \cos(f_k u_1)\cos(f_k u_2) \big] \\
     &= \sum_{k=0}^{K-1} \cos\big(f_k (u_1 - u_2)\big),
    \end{aligned}
\end{equation}
which depends only on the difference $u_1 - u_2$ rather than on absolute positions. Moreover, since $f_k \le f_0 = 1$, every term $\cos(f_k \Delta)$ is decreasing in $\Delta$ on $[0, \pi]$, so the inner product is strictly decreasing in $|u_1 - u_2|$ whenever $|u_1 - u_2| \le \pi$, a range that covers all intra-city displacements by a wide margin. Embedding similarity therefore behaves as a monotonically decreasing function of coordinate displacement. Note that although longitude is periodic ($-180^\circ \equiv 180^\circ$), all POIs in our setting lie within a single city, so coordinate displacements satisfy $|u_1 - u_2| \ll \pi$ and the wrap-around case does not arise in practice.

\textbf{Property 2 (Lipschitz continuity).} Expanding the squared distance between two feature vectors,
\begin{equation}
    \begin{aligned}
    \| \gamma(u_1) - \gamma(u_2) \|_2^2 
    &= \sum_{k=0}^{K-1} 2\big(1 - \cos(f_k (u_1 - u_2))\big) \\
    &\;\le\; \Big( \sum_{k=0}^{K-1} f_k^2 \Big) (u_1 - u_2)^2,
    \end{aligned}
\end{equation}
where the inequality follows from $1 - \cos\theta \le \theta^2/2$. The encoding is thus Lipschitz-continuous: nearby coordinates are guaranteed nearby features. The subsequent MLP, a composition of Lipschitz maps (linear layers and ReLU), preserves this continuity, so POIs with adjacent coordinates enter RQ-Kmeans with adjacent geospatial embeddings $\mathbf{e}_{\text{geo}}$, the prerequisite for adjacent SI-SIDs. The frequencies $f_k = 10000^{-k/K}$ are exponentially spaced, so different bands respond to coordinate differences at different rates, and the MLP learns a weighted combination across bands during end-to-end training.

\textbf{Remark.} Both properties are stated for the Fourier features $\gamma$. The MLP preserves continuity exactly (Property 2), whereas the distance--similarity monotonicity of Property 1 is preserved qualitatively after the learned projection and verified empirically (Figure~\ref{fig:geo_property}). In addition, the kernel depends on differences of normalized coordinates rather than geodesic distance; within a single city the two are approximately proportional, so the stated monotonicity carries over to real distances at the scale relevant to next-POI recommendation.

\subsection{Empirical Verification}

We further verify that the two exact properties of $\gamma$ survive the MLP projection, i.e., that they hold for the final geospatial embedding $\mathbf{e}_{\text{geo}}(\mathbf{p})$ of a location $\mathbf{p} = (\lambda, \phi)$. To this end, we encode a $30\times30$ uniform coordinate grid over an urban region of Beijing ($116.2^\circ$--$116.6^\circ$E, $39.8^\circ$--$40.1^\circ$N; 900 points covering roughly $44\,\text{km}\times33\,\text{km}$, Figure~\ref{fig:geo_property}(a)) and inspect the geometry of the resulting embeddings, as summarized in Figure~\ref{fig:geo_property}.

First, shift-invariance manifests itself in the regularity of the embedded lattice. As shown in Figure~\ref{fig:geo_property}(b), the t-SNE projection of the 900 embeddings still forms an evenly spaced grid that mirrors the original one: the same coordinate step induces a nearly identical embedding change wherever it is applied, and no region of the city is stretched or compressed. This position-independent response is exactly the behavior predicted by Property~1.

Second, continuity and metric preservation are confirmed by the distance relation in Figure~\ref{fig:geo_property}(c): over sampled point pairs of the grid, the embedding distance grows monotonically and near-linearly with the geographic distance (Pearson $r = 0.893$). Nearby coordinates thus receive nearby embeddings, while farther places are consistently mapped farther apart, matching Property~2.

Together, these results confirm that POIs enter SI-SID construction with a geometry-faithful geospatial embedding. Notably, this encoder-level property is corroborated by the identifier-level structure observed in the main text: because the fused representation carries an explicit metric structure over coordinates, the first-level quantization inherits it and partitions POIs primarily by geographic proximity, with the top first-level clusters occupying compact and mutually distinct urban regions. Once the first-level codeword absorbs this shared geographic component, the residual passed to deeper levels is dominated by semantic variation. Accordingly, the second-level sub-clusters within a single first-level cluster overlap geographically yet separate cleanly by semantic category. The geometry verified here at the encoder level and the hierarchy observed at the identifier level thus validate each other, closing the chain from coordinate encoding to spatially continuous identifiers.

\section{Details of the MG-CPT Training Corpus}
\label{app:mgcpt_data}

The goal of MG-CPT is to internalize the spatial relations of the whole city into the LLM, so that POIs become interconnected places rather than isolated tokens. To this end, we build a training corpus that spans points, lines, and flows over urban space, comprising \textbf{25 datasets}, one per task type, each serialized into \texttt{(instruction, input, output)} triples. Specifically, the \texttt{instruction} assigns a city-geography expert role and states the requirement, the \texttt{input} is the question, and the \texttt{output} is the answer. For all open-ended reasoning tasks, the \texttt{output} follows an ``answer-first, derivation-second'' format written in plain language, without any special reasoning tags. The corpus is derived from four kinds of geospatial data, summarized in Table~\ref{tab:mgcpt_sources}.

\begin{table*}[t]
    \centering
    \caption{The four data sources of the MG-CPT training corpus.}
    \label{tab:mgcpt_sources}
    \scalebox{0.9}{
    \begin{tabular}{l|l|c}
    \hline
    \hline
    Source & Information provided & Granularity \\
    \hline
    POI snapshot & SI-SID, name, coordinates, address, category, district, \ldots & point (static) \\
    Road feature & road class, length, chain/link count, lanes, speed limit, \ldots & line (static) \\
    POI-pair relation & distance and direction of a POI pair, with derivations, \ldots & point--point \\
    Navigation route & driving distance, duration, road sequence, turn-by-turn steps, \ldots & point--line--point (dynamic) \\
    \hline
    \hline
    \end{tabular}
    }
\end{table*}

\begin{figure*}[t]
    \centering
                \subfloat[Mainly static road network rendered from road features (Beijing).]{
         \begin{minipage}[t]{0.5\linewidth}
         \centering
         \includegraphics[width=0.8\textwidth]{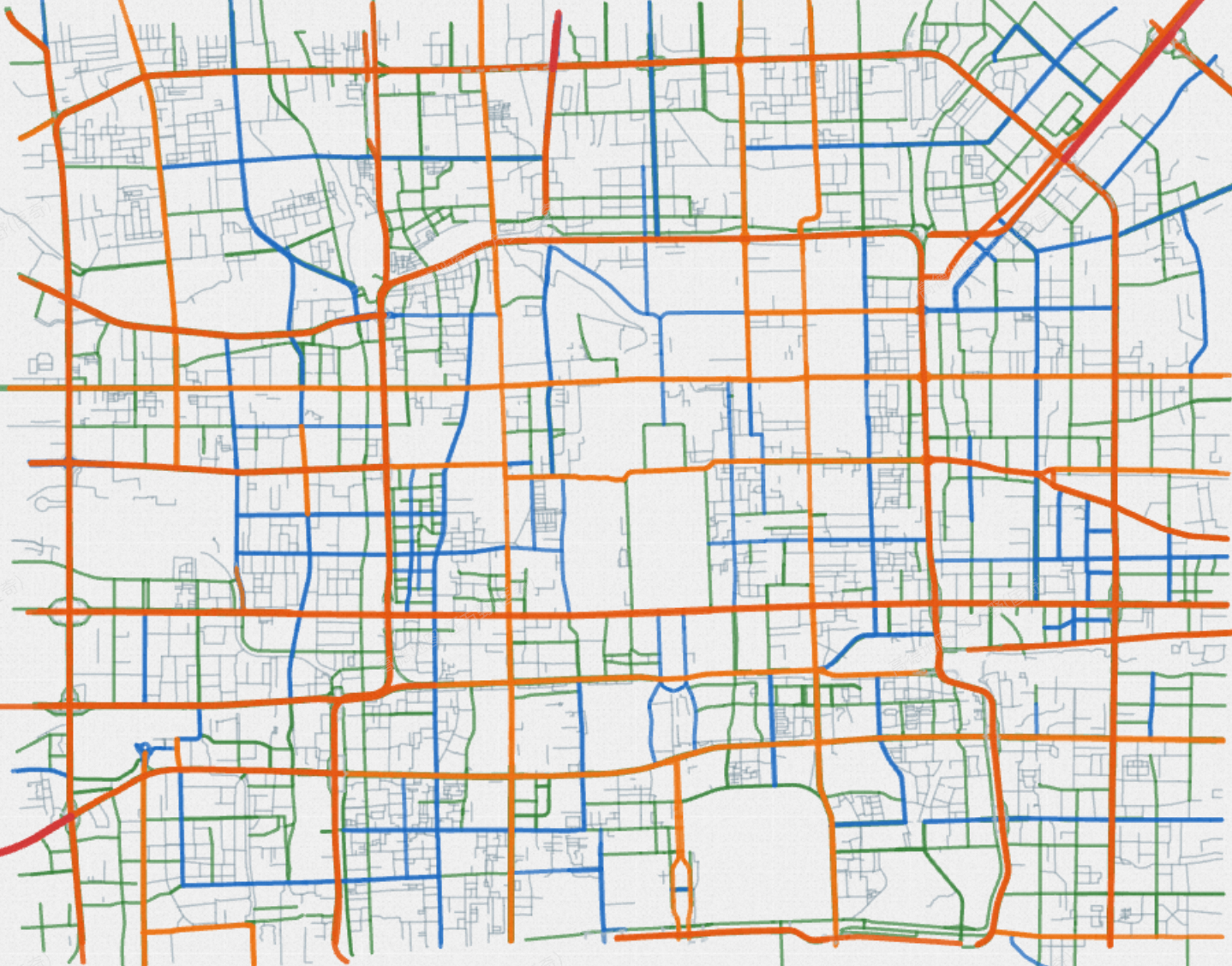}
         \end{minipage}
        } 
        \subfloat[Dynamic flow aggregated from 2{,}000 navigation trajectories; each trajectory is a POI--road--POI connection.]{
         \begin{minipage}[t]{0.5\linewidth}
         \centering
         \includegraphics[width=0.74\textwidth]{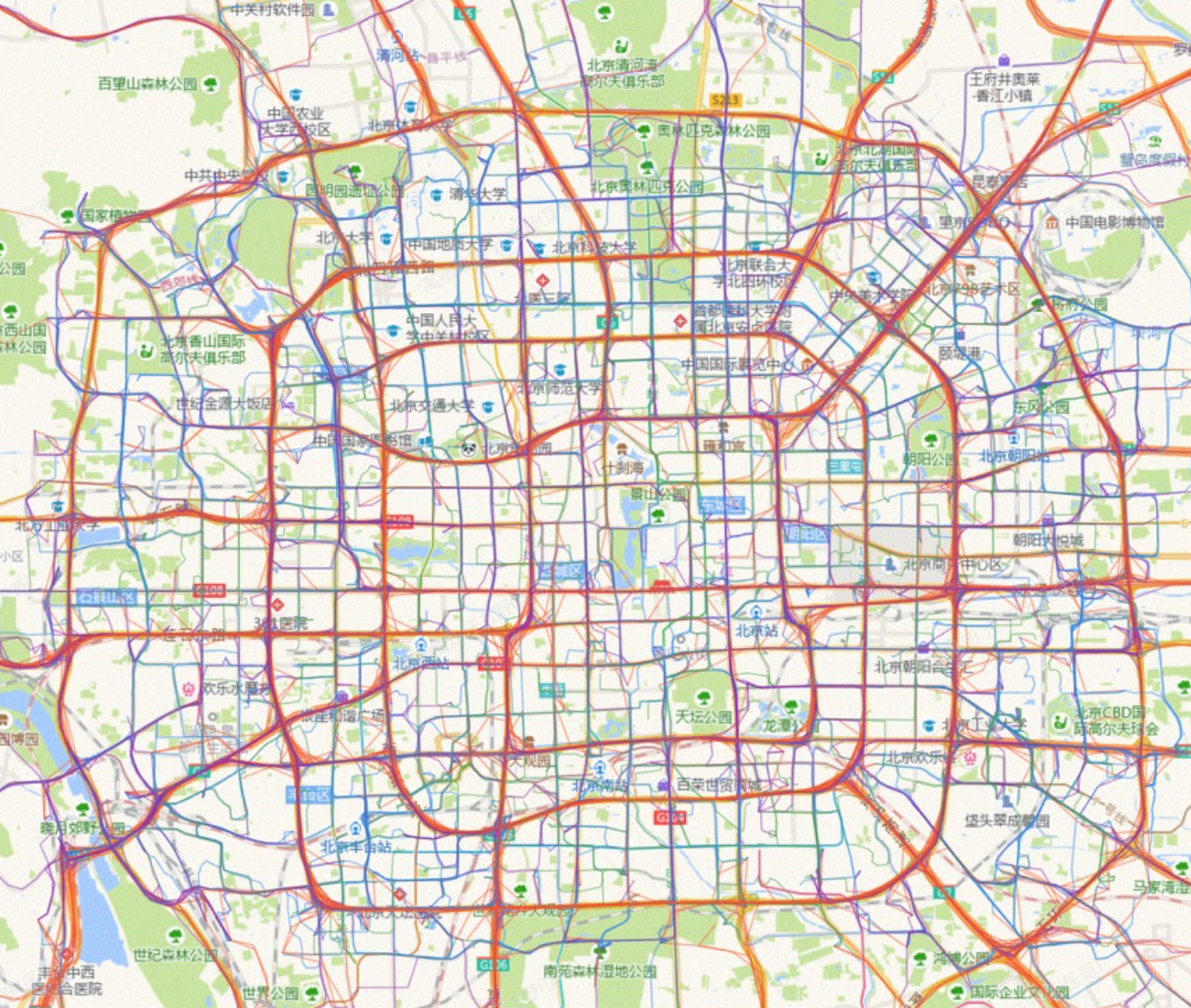}
         \end{minipage}
        }
    \caption{Two complementary views of Beijing: the static road skeleton and the dynamic movement flow.}
\label{fig:city_map}
\end{figure*}

\subsection{Three Tiers of Datasets}

The 25 training datasets instantiate the three tiers of MG-CPT and together assemble a virtual city street map that the backbone can internalize. \textbf{Basic-attribute (\texttt{B}, 13 datasets)} describe the static attributes of a single object: twelve POI datasets establish the bidirectional mapping among a POI's SI-SID, coordinates, and attributes (category, district, address), while one road dataset characterizes a single road (class, length, lanes, speed limit). This tier is the foundation on which all other spatial abilities rest. \textbf{Relational (\texttt{R}, 9 datasets)} target pairwise geometry; the eight distance/direction datasets form a fully orthogonal $2\times2\times2$ design over coordinate visibility (\texttt{coord}/\texttt{nocoord}), reasoning target (\texttt{dis}/\texttt{dir}), and answer format (\texttt{choice}/\texttt{open}), and the contrast between the \texttt{coord} and \texttt{nocoord} groups directly quantifies how much positional knowledge has been internalized, since the latter provides no numeric hint. \textbf{Systemic (\texttt{S}, 3 datasets)} move from points to the road network, connecting an origin, a path, and a destination into a complete movement through three progressively finer questions: how far, which roads, and how to drive step by step.

\subsection{Static Road Network and Dynamic Flow}

Internalizing the spatial relations of the whole city relies on two complementary signals: the static road skeleton and the dynamic flow of movement. The road-feature tasks capture the static skeleton---the class, orientation, and connectivity of arterials, ring roads, and expressways---which alone outlines the spatial framework of a city, as visualized in Figure~\ref{fig:city_map}(a). However, the static skeleton does not reveal how people actually move through the city. Navigation data encode real movement in the form of an origin POI, a traversed road sequence, and a destination POI, so that each route is a POI--road--POI connection. As shown in Figure~\ref{fig:city_map}(b), even 2{,}000 navigation trajectories (the training set uses about 50{,}000) already reveal the main flow arteries of the city; when a large number of such connections are aggregated, isolated POIs are woven by real travel paths into a dynamic virtual street map of Beijing. The contrast between the two figures makes the value of this design concrete. Figure~\ref{fig:city_map}(a) alone tells the model where the roads are, but the streets remain empty: it cannot say which places people actually travel between, or by what route. Figure~\ref{fig:city_map}(b) fills these streets with real movement: every trajectory threads an origin POI, a sequence of roads, and a destination POI into a single line, and as thousands of them overlay, the busy corridors of the city light up while previously isolated POIs turn into nodes on a living transportation graph. By continually pre-training on both views, MG-CPT internalizes not only the straight-line distance between two POIs, but also how one is reached from another through the road network. This directly benefits downstream next-POI recommendation: rather than merely fitting behavioral co-occurrence, the model can favor destinations that are genuinely reachable from the user's current location, so that its recommendations respect real urban geography instead of historical transition patterns alone.

\subsection{Distance and Direction Annotation}

\textbf{Spherical distance (Haversine).} Open-ended distance tasks provide a full derivation. For two points $(\lambda_1,\varphi_1)$ and $(\lambda_2,\varphi_2)$,
\begin{equation}
    a=\sin^2\!\Big(\tfrac{\Delta\varphi}{2}\Big)+\cos\varphi_1\cos\varphi_2\sin^2\!\Big(\tfrac{\Delta\lambda}{2}\Big),\quad d=2R\arcsin\sqrt{a},
\end{equation}
with $R=6371$~km. \textbf{Planar approximation.} For direction judgment and some tasks we adopt a city-scale planar approximation $d\approx\sqrt{(\Delta\lambda\cdot 85)^2+(\Delta\varphi\cdot 111)^2}$~km, where $111$ is the kilometers per degree of latitude and $85$ that per degree of longitude at the city's latitude. \textbf{Eight-way direction.} The sign of $\Delta\lambda$ and $\Delta\varphi$ determines east/west and north/south, and comparing the east--west displacement $|\Delta\lambda|\cdot 85$ with the north--south displacement $|\Delta\varphi|\cdot 111$ decides between a principal and a composite direction, over the set \{N, NE, E, SE, S, SW, W, NW\}.

\subsection{Option and Distractor Construction}


The position of the correct option is determined by a hash of the primary key so that A--D are uniformly distributed and reproducible. Numeric distractors scale the ground truth ($\times1.5$, $\times0.6$, $\times2.0$, $\times0.4$ for distances) or progressively offset the coordinates, while categorical distractors are drawn from a circular list with offsets coprime to its length (e.g., eight directions or sixteen districts), which prevents duplicated or trivially eliminable options. Open-ended distance and direction tasks additionally include explicit derivations, and the three POI-description tasks each traverse the full POI set, differing not in which POIs they cover but in which attributes serve as the question and which as the answer.

\begin{table*}[t]
    \setlength{\abovecaptionskip}{0.2cm}
    \setlength{\belowcaptionskip}{-0.4cm}
    \centering
    \caption{The 25 training datasets of the MG-CPT training corpus (from the Shanghai dataset). ``Num(POI)'' and ``Num(Road)'' denote the numbers of POIs and roads, respectively. Additionally, ``MC'' and ``Open'' denote the multiple-choice and open-ended template types. For brevity, the "Input $\rightarrow$ Output'' column lists only representative attributes rather than the full set involved in each dataset. }
    \label{tab:mgcpt_corpus}
    \scalebox{0.9}{
    \begin{tabular}{c|l|c|p{10cm}|l}
    \hline
    \hline
    Tier & Task type & Format & Input $\rightarrow$ Output & Scale \\
    \hline
    \multirow{13}{*}{\shortstack{Basic\\(\texttt{B})}}
     & \texttt{poi\_coord\_qa}        & Open   & SI-SID $\rightarrow$ coordinates & Num(POI) \\
     & \texttt{poi\_coord\_reverse}   & Open   & coordinates $\rightarrow$ SI-SID, name, district & Num(POI) \\
     & \texttt{poi\_describe\_name}   & Open   & name, SI-SID $\rightarrow$ category, district, address, coordinates & Num(POI) \\
     & \texttt{poi\_describe\_coord}  & Open   & coordinates $\rightarrow$ SI-SID, category, district, address & Num(POI) \\
     & \texttt{poi\_describe\_reverse}& Open   & name, district, address, category $\rightarrow$ SI-SID, coordinates & Num(POI) \\
     \cline{2-5}
     & \texttt{qa\_category}          & Open   & SI-SID $\rightarrow$ category & 20,000 \\
     & \texttt{qa\_coord\_choice}     & MC & SI-SID + 4 coordinate options $\rightarrow$ option & 20,000 \\
     & \texttt{qa\_coord\_open}       & Open   & SI-SID $\rightarrow$ coordinates & 20,000 \\
     & \texttt{qa\_coord\_open\_full} & Open   & category + district + address $\rightarrow$ coordinates, SI-SID & 20,000 \\
     & \texttt{qa\_district\_choice}  & MC & SI-SID + 4 district options $\rightarrow$ option & 20,000 \\
     & \texttt{qa\_district\_open}    & Open   & SI-SID $\rightarrow$ district & 20,000 \\
     & \texttt{qa\_nearby}            & Open   & coordinates $\rightarrow$ SI-SID, category, district, address & 20,000 \\
     & \texttt{road\_describe}        & Open   & road name $\rightarrow$ class, length, chain/link count, lanes, speed limit & Num(Road) \\
    \hline
    \multirow{9}{*}{\shortstack{Relational\\(\texttt{R})}}
     & \texttt{poi\_nearby\_to\_poi}  & MC & reference coordinates + 4 candidates $\rightarrow$ nearest option (with distances) & 20,000 \\
     & \texttt{dis\_coord\_choice}    & MC & two SI-SIDs + coordinates $\rightarrow$ distance option & 20,000 \\
     & \texttt{dis\_coord\_open}      & Open   & two SI-SIDs + coordinates $\rightarrow$ distance, with Haversine derivation & 20,000 \\
     & \texttt{dir\_coord\_choice}    & MC & two SI-SIDs + coordinates $\rightarrow$ direction option & 20,000 \\
     & \texttt{dir\_coord\_open}      & Open   & two SI-SIDs + coordinates $\rightarrow$ direction, with derivation & 20,000 \\
     & \texttt{dis\_nocoord\_choice}  & MC & two SI-SIDs only $\rightarrow$ distance option & 20,000 \\
     & \texttt{dis\_nocoord\_open}    & Open   & two SI-SIDs only $\rightarrow$ distance, recalling coordinates first & 20,000 \\
     & \texttt{dir\_nocoord\_choice}  & MC & two SI-SIDs only $\rightarrow$ direction option & 20,000 \\
     & \texttt{dir\_nocoord\_open}    & Open   & two SI-SIDs only $\rightarrow$ direction, recalling coordinates first & 20,000 \\
    \hline
    \multirow{3}{*}{\shortstack{Systemic\\(\texttt{S})}}
     & \texttt{route\_dist}           & Open   & origin, destination $\rightarrow$ driving distance, duration, districts & 50,000 \\
     & \texttt{route\_roads}          & Open   & origin, destination $\rightarrow$ distance, duration, road sequence & 50,000 \\
     & \texttt{route\_navi}           & Open   & origin, destination $\rightarrow$ turn-by-turn navigation & 50,000 \\
    \hline
    \hline
    \end{tabular}
    }
\end{table*}

Table~\ref{tab:mgcpt_corpus} lists the templates of all 25 training tasks. To avoid leaking proprietary data, in the following demonstrations we denote a POI's SI-SID by the placeholder \texttt{\{SI-SID$_i$\}} and a coordinate by $(lon_i, lat_i)$; other public information (names, addresses, road names) is kept for readability.

\subsection{Dataset Demonstrations}

We show a full card for every one of the 25 training tasks, each consisting of the (translated) instruction, the question, and the gold answer. To avoid leaking proprietary data, SI-SIDs, coordinates, POI names, addresses, and road names are all replaced by placeholders, \texttt{\{SI-SID$_i$\}}, $(lon_i,lat_i)$, \texttt{\{name$_i$\}}, \texttt{\{addr$_i$\}}, and \texttt{\{road$_i$\}}, respectively, while generic attributes such as category and district are kept for readability.

\subsubsection{Basic-attribute: POI point (12 tasks).}
\begin{itemize}[leftmargin=1.2em]
    \item \textbf{\texttt{poi\_coord\_qa}} \\
    \textit{Instruction:} You are a geography expert of Shanghai, familiar with the coordinates, districts, and addresses of all local POIs. Accurately provide the coordinates and district of the given POI. \\
    \textit{Question:} What are the coordinates of \texttt{\{SI-SID$_0$\}}? \\
    \textit{Answer:} $(lon_0, lat_0)$.
    \item \textbf{\texttt{poi\_coord\_reverse}} \\
    \textit{Instruction:} You are a geography expert of Shanghai. Given a coordinate, accurately identify the corresponding POI, its name, and its district. \\
    \textit{Question:} Which POI is located at $(lon_0, lat_0)$? \\
    \textit{Answer:} \texttt{\{name$_0$\}} (\texttt{\{SI-SID$_0$\}}), in Huangpu District.
    \item \textbf{\texttt{poi\_describe\_name}} \\
    \textit{Instruction:} You are a geography expert of Shanghai. Given a POI's name or SID, accurately provide its coordinates, district, and address. \\
    \textit{Question:} Describe the basic geographic information of \texttt{\{SI-SID$_0$\}}. \\
    \textit{Answer:} \texttt{\{SI-SID$_0$\}} is a shopping mall in Huangpu District, at \texttt{\{addr$_0$\}}, with coordinates $(lon_0, lat_0)$.
    \item \textbf{\texttt{poi\_describe\_coord}} \\
    \textit{Instruction:} You are a geography expert of Shanghai. Given a coordinate, provide the POI located there together with its category, district, and address. \\
    \textit{Question:} What is located at $(lon_0, lat_0)$? \\
    \textit{Answer:} \texttt{\{SI-SID$_0$\}}, a shopping mall in Huangpu District, at \texttt{\{addr$_0$\}}.
    \item \textbf{\texttt{poi\_describe\_reverse}} \\
    \textit{Instruction:} You are a geography expert of Shanghai. Given address information, accurately provide the corresponding POI name, coordinates, and district. \\
    \textit{Question:} Which POI is a shopping mall at \texttt{\{addr$_0$\}}, Huangpu District? \\
    \textit{Answer:} This is \texttt{\{name$_0$\}} (\texttt{\{SI-SID$_0$\}}), with coordinates $(lon_0, lat_0)$.
    \item \textbf{\texttt{qa\_category}} \\
    \textit{Instruction:} You are a geography expert of Shanghai. Answer the question concisely and accurately. \\
    \textit{Question:} What is the category of \texttt{\{SI-SID$_0$\}}? \\
    \textit{Answer:} Shopping mall.
    \item \textbf{\texttt{qa\_coord\_choice}} \\
    \textit{Instruction:} You are a geography expert of Shanghai. For multiple-choice questions, directly output the correct option letter. \\
    \textit{Question:} What are the coordinates of \texttt{\{SI-SID$_0$\}}? A.~$(lon_1,lat_1)$ B.~$(lon_2,lat_2)$ C.~$(lon_3,lat_3)$ D.~$(lon_4,lat_4)$ \\
    \textit{Answer:} B.
    \item \textbf{\texttt{qa\_coord\_open}} \\
    \textit{Instruction:} You are a geography expert of Shanghai. Answer the question concisely and accurately. \\
    \textit{Question:} Give the coordinates of \texttt{\{SI-SID$_0$\}}. \\
    \textit{Answer:} $(lon_0, lat_0)$.
    \item \textbf{\texttt{qa\_coord\_open\_full}} \\
    \textit{Instruction:} You are a geography expert of Shanghai. Given the category, district, and address, provide the coordinates and SI-SID of the POI. \\
    \textit{Question:} Where is the shopping mall at \texttt{\{addr$_0$\}}, Huangpu District? \\
    \textit{Answer:} $(lon_0, lat_0)$, i.e., \texttt{\{SI-SID$_0$\}}.
    \item \textbf{\texttt{qa\_district\_choice}} \\
    \textit{Instruction:} You are a geography expert of Shanghai. For multiple-choice questions, directly output the correct option letter. \\
    \textit{Question:} Which district of Shanghai is \texttt{\{SI-SID$_0$\}} in? A.~Songjiang B.~Chongming C.~Huangpu D.~Jiading \\
    \textit{Answer:} C.
    \item \textbf{\texttt{qa\_district\_open}} \\
    \textit{Instruction:} You are a geography expert of Shanghai. Answer the question concisely and accurately. \\
    \textit{Question:} Which district of Shanghai is \texttt{\{SI-SID$_0$\}} in? \\
    \textit{Answer:} Huangpu District.
    \item \textbf{\texttt{qa\_nearby}} \\
    \textit{Instruction:} You are a geography expert of Shanghai. Answer the question concisely and accurately. \\
    \textit{Question:} Which POI is near the coordinate $(lon_0, lat_0)$? \\
    \textit{Answer:} Near this coordinate is \texttt{\{SI-SID$_0$\}}, a shopping mall in Huangpu District, at \texttt{\{addr$_0$\}}.
\end{itemize}

\subsubsection{Basic-attribute: line (1 task).}
\begin{itemize}[leftmargin=1.2em]
    \item \textbf{\texttt{road\_describe}} \\
    \textit{Instruction:} You are a geography expert of Beijing, familiar with the detailed attributes of the city's road network. Accurately describe the class, length, lanes, and speed limit of the given road. \\
    \textit{Question:} Describe the basic road information of \texttt{\{road$_0$\}}, Beijing. \\
    \textit{Answer:} \texttt{\{road$_0$\}} is a secondary road about 31.4 km long, consisting of 26 directional chains and 361 links, mostly 6 lanes, with a speed limit of 70 km/h.
\end{itemize}

\subsubsection{Relational: point to point (9 tasks).}
\begin{itemize}[leftmargin=1.2em]
    \item \textbf{\texttt{poi\_nearby\_to\_poi}} \\
    \textit{Instruction:} You are a geography expert of Shanghai. For multiple-choice questions, directly output the correct option letter. \\
    \textit{Question:} Which of the following coordinates is closest to $(lon_0, lat_0)$? A.~$(lon_1,lat_1)$ B.~$(lon_2,lat_2)$ C.~$(lon_3,lat_3)$ D.~$(lon_4,lat_4)$ \\
    \textit{Answer:} B. (The answer computes the distance to each option).
    \item \textbf{\texttt{dis\_coord\_choice}} \\
    \textit{Instruction:} You are a geography expert. Based on the two coordinates, judge the straight-line distance. For multiple-choice questions, output the option letter. \\
    \textit{Question:} Distance between \texttt{\{SI-SID$_0$\}} $(lon_0,lat_0)$ and \texttt{\{SI-SID$_1$\}} $(lon_1,lat_1)$? A.~about 4.7 km B.~about 7.9 km C.~about 15.8 km D.~about 3.2 km \\
    \textit{Answer:} B.
    \item \textbf{\texttt{dis\_coord\_open}} \\
    \textit{Instruction:} You are a geography expert. Based on the two coordinates, compute the straight-line distance; give the answer first, then the derivation. \\
    \textit{Question:} What is the straight-line distance between \texttt{\{SI-SID$_0$\}} $(lon_0,lat_0)$ and \texttt{\{SI-SID$_1$\}} $(lon_1,lat_1)$? \\
    \textit{Answer:} About 7.9 km. \{Followed by the Haversine derivation.\}
    \item \textbf{\texttt{dir\_coord\_choice}} \\
    \textit{Instruction:} You are a geography expert. Based on the two coordinates, judge the direction of A relative to B. For multiple-choice questions, output the option letter. \\
    \textit{Question:} Direction of \texttt{\{SI-SID$_0$\}} $(lon_0,lat_0)$ relative to \texttt{\{SI-SID$_1$\}} $(lon_1,lat_1)$? A.~Northeast B.~Southeast C.~Southwest D.~Northwest \\
    \textit{Answer:} C.
    \item \textbf{\texttt{dir\_coord\_open}} \\
    \textit{Instruction:} You are a geography expert. Based on the two coordinates, reason about the direction of A relative to B; give the answer first, then the reasoning. \\
    \textit{Question:} Reason about the direction of \texttt{\{SI-SID$_0$\}} $(lon_0,lat_0)$ relative to \texttt{\{SI-SID$_1$\}} $(lon_1,lat_1)$. \\
    \textit{Answer:} Southwest. \{With a component-wise analysis.\}
    \item \textbf{\texttt{dis\_nocoord\_choice}} \\
    \textit{Instruction:} You are a geography expert. Based only on the two POIs' SIDs, judge the straight-line distance. For multiple-choice questions, output the option letter. \\
    \textit{Question:} Distance between \texttt{\{SI-SID$_0$\}} and \texttt{\{SI-SID$_1$\}}? A.~about 4.7 km B.~about 7.9 km C.~about 15.8 km D.~about 3.2 km \\
    \textit{Answer:} B.
    \item \textbf{\texttt{dis\_nocoord\_open}} \\
    \textit{Instruction:} You are a geography expert. Based only on the two POIs' SIDs, judge the straight-line distance; give the answer first, then the analysis. \\
    \textit{Question:} Approximately how far apart are \texttt{\{SI-SID$_0$\}} and \texttt{\{SI-SID$_1$\}}? \\
    \textit{Answer:} About 7.9 km; the answer first recalls the two coordinates, then derives the distance.
    \item \textbf{\texttt{dir\_nocoord\_choice}} \\
    \textit{Instruction:} You are a geography expert. Based only on the two POIs' SIDs, judge the direction. For multiple-choice questions, output the option letter. \\
    \textit{Question:} Direction of \texttt{\{SI-SID$_0$\}} relative to \texttt{\{SI-SID$_1$\}}? A.~Northeast B.~Southeast C.~Southwest D.~Northwest \\
    \textit{Answer:} C.
    \item \textbf{\texttt{dir\_nocoord\_open}} \\
    \textit{Instruction:} You are a geography expert. Based only on the two POIs' SIDs, reason about the direction; give the answer first, then the analysis. \\
    \textit{Question:} Reason about the direction of \texttt{\{SI-SID$_0$\}} relative to \texttt{\{SI-SID$_1$\}}. \\
    \textit{Answer:} Southwest; the answer recalls the coordinates, then analyzes the components.
\end{itemize}

\subsubsection{Systemic: point--line--point, navigation (3 tasks).}
\begin{itemize}[leftmargin=1.2em]
    \item \textbf{\texttt{route\_dist}} \\
    \textit{Instruction:} You are a geography expert of Shanghai. From the origin and destination POI SIDs, you can obtain approximate navigation information, including distance and time. \\
    \textit{Question:} How far is it from \texttt{\{SI-SID$_0$\}} to \texttt{\{SI-SID$_1$\}}, and how long does it take? \\
    \textit{Answer:} About 11.3 km, roughly 28 min by car (a road-network distance, distinct from the 7.9 km straight-line distance).
    \item \textbf{\texttt{route\_roads}} \\
    \textit{Instruction:} You are a geography expert of Shanghai. Based on the origin and destination SIDs, list the main roads passed by this trip. \\
    \textit{Question:} Which roads should one take from \texttt{\{SI-SID$_0$\}} to \texttt{\{SI-SID$_1$\}}? \\
    \textit{Answer:} About 11.3 km, roughly 28 min, mainly via \texttt{\{road$_1$\}}; \texttt{\{road$_2$\}}; \texttt{\{road$_3$\}}; \ldots
    \item \textbf{\texttt{route\_navi}} \\
    \textit{Instruction:} You are a geography expert of Shanghai. From the origin and destination POI SIDs, you can obtain detailed navigation. \\
    \textit{Question:} Give the detailed navigation from \texttt{\{SI-SID$_0$\}} to \texttt{\{SI-SID$_1$\}}. \\
    \textit{Answer:} Head north on \texttt{\{road$_1$\}} for 180 m, turn right onto \texttt{\{road$_2$\}}; \ldots; arrive at the destination.
\end{itemize}

\begin{table*}[t]
    \setlength{\abovecaptionskip}{0.2cm}
    \setlength{\belowcaptionskip}{-0.4cm}
    \centering
    \caption{Details and demonstrations of the 18 spatial cognition evaluation tasks for MG-CPT (examples from the Shanghai dataset). \texttt{B}/\texttt{R}/\texttt{S} denote Basic-attribute, Relational, and Systemic categories; \texttt{coord}/\texttt{no\_coord} indicates whether coordinates are provided. \texttt{\{SI-SID$_i$\}} is a placeholder for the four-token SI-SID of a POI. For brevity, the ``Example'' column shows abridged questions, with long navigation steps and road sequences truncated by ``\ldots''; the system prompt prepended to each question at evaluation time is omitted.}
    \label{tab:mgcpt_tasks}
    \scalebox{0.82}{
    \begin{tabular}{c|l|p{4.2cm}|p{9.9cm}|c}
    \hline
    \hline
    Format & Task & Description & Example & Answer \\
    \hline
    \multirow{38}{*}{MC}
     & \texttt{B\_poi\_coord} & Identify the coordinates of a given POI. & What are the coordinates of \texttt{\{SI-SID$_0$\}}? A.~$(lon_1,lat_1)$ B.~$(lon_2,lat_2)$ C.~$(lon_3,lat_3)$ D.~$(lon_4,lat_4)$ & A \\
     \cline{2-5}
     & \texttt{B\_poi\_area\_no\_coord} & Identify the district a POI belongs to, without coordinates. & Which district of Shanghai is \texttt{\{SI-SID$_0$\}} located in? A.~Qingpu B.~Jiading C.~Xuhui D.~Hongkou & B \\
      \cline{2-5}
     & \texttt{B\_poi\_area\_coord} & Identify the district a POI belongs to, with coordinates. & Which district of Shanghai is \texttt{\{SI-SID$_0$\}} $(lon_0,lat_0)$ located in? A.~Qingpu B.~Jiading C.~Xuhui D.~Hongkou & B \\
      \cline{2-5}
     & \texttt{R\_poi\_nearby\_poi} & Select the POI nearest to a given POI. & Which POI is closest to the coordinate $(lon_0,lat_0)$? A.~\texttt{\{SI-SID$_0$\}} B.~\texttt{\{SI-SID$_1$\}} C.~\texttt{\{SI-SID$_2$\}} D.~\texttt{\{SI-SID$_3$\}} & D \\
      \cline{2-5}
     & \texttt{R\_poi\_nearby\_coord} & Select the coordinate nearest to a given coordinate. & Which of the following coordinates is nearest to $(lon_0,lat_0)$? A.~$(lon_1,lat_1)$ B.~$(lon_2,lat_2)$ C.~$(lon_3,lat_3)$ D.~$(lon_4,lat_4)$ & C \\
      \cline{2-5}
     & \texttt{R\_poi\_dis\_no\_coord} & Estimate the distance between two POIs, without coordinates. & What is the approximate distance between \texttt{\{SI-SID$_0$\}} and \texttt{\{SI-SID$_1$\}}? A.~about 18.9~km B.~about 37.8~km C.~about 75.6~km D.~about 113.4~km & B \\
      \cline{2-5}
     & \texttt{R\_poi\_dis\_coord} & Estimate the distance between two POIs, with coordinates. & What is the approximate distance between \texttt{\{SI-SID$_0$\}} $(lon_0,lat_0)$ and \texttt{\{SI-SID$_1$\}} $(lon_1,lat_1)$? A.~about 18.9~km B.~about 37.8~km C.~about 75.6~km D.~about 113.4~km & B \\
      \cline{2-5}
     & \texttt{R\_poi\_dir\_no\_coord} & Judge the direction between two POIs, without coordinates. & In which direction is \texttt{\{SI-SID$_0$\}} relative to \texttt{\{SI-SID$_1$\}}? A.~North B.~East C.~West D.~South & D \\
      \cline{2-5}
     & \texttt{R\_poi\_dir\_coord} & Judge the direction between two POIs, with coordinates. & In which direction is \texttt{\{SI-SID$_0$\}} $(lon_0,lat_0)$ relative to \texttt{\{SI-SID$_1$\}} $(lon_1,lat_1)$? A.~North B.~East C.~West D.~South & D \\
     & \texttt{S\_navi\_detail\_dest} & Infer the destination from a detailed step-by-step navigation route. & Starting from \texttt{\{SI-SID$_0$\}}$(lon_0,lat_0)$, follow the detailed navigation: 1.~drive west for 55~m and turn right; 2.~drive north for 21~m and turn right onto the main road; \ldots; 20.~along the pedestrian street for 119~m to the destination (37.2~km, 68~min in total). Which POI is the most likely destination? A.~\texttt{\{SI-SID$_1$\}} B.~\texttt{\{SI-SID$_2$\}} C.~\texttt{\{SI-SID$_3$\}} D.~\texttt{\{SI-SID$_4$\}} & B \\
      \cline{2-5}
     & \texttt{S\_navi\_via\_poi} & Judge which stopover POI is more on the way of a trip. & I depart from Zixing Tobacco Store for Gongyuan Pipeline and want to stop by either the waste recycling station (22.6~km, 43~min) or the Linjiang Road Ferry Terminal (23.7~km, 61~min). Which stopover is more on the way? A.~via the recycling station B.~via the ferry terminal C.~equally convenient D.~cannot be determined & A \\
      \cline{2-5}
     & \texttt{S\_navi\_simple\_dest} & Infer the destination from a road-level navigation route. & Starting from \texttt{\{SI-SID$_0$\}} $(lon_0,lat_0)$, drive along Longqi Rd; Longteng Ave; Xumei Rd; Humin Interchange; G60 Expressway exit; Jiamin Elevated Rd; \ldots; Pedestrian St (37.2~km, 68~min in total). Which POI is the most likely destination? A.~\texttt{\{SI-SID$_1$\}} B.~\texttt{\{SI-SID$_2$\}} C.~\texttt{\{SI-SID$_3$\}} D.~\texttt{\{SI-SID$_4$\}} & B \\
    \hline
    \multirow{12}{*}{Open}
     & \texttt{B\_poi\_area\_no\_coord} & State the district a POI belongs to, without coordinates. & Which district of Shanghai is \texttt{\{SI-SID$_0$\}} located in? & Jiading \\
      \cline{2-5}
     & \texttt{B\_poi\_area\_coord} & State the district a POI belongs to, with coordinates. & Which district of Shanghai is \texttt{\{SI-SID$_0$\}} $(lon_0,lat_0)$ located in? & Jiading \\
      \cline{2-5}
     & \texttt{R\_poi\_dir\_no\_coord} & State the direction between two POIs, without coordinates. & In which direction is \texttt{\{SI-SID$_0$\}} relative to \texttt{\{SI-SID$_1$\}}? & East \\
      \cline{2-5}
     & \texttt{R\_poi\_dir\_coord} & State the direction between two POIs, with coordinates. & In which direction is \texttt{\{SI-SID$_0$\}} $(lon_0,lat_0)$ relative to \texttt{\{SI-SID$_1$\}} $(lon_1,lat_1)$? & East \\
      \cline{2-5}
     & \texttt{R\_poi\_dis\_no\_coord} & Estimate the distance between two POIs, without coordinates. & What is the approximate distance between \texttt{\{SI-SID$_0$\}} and \texttt{\{SI-SID$_1$\}}? & 37.8~km \\
      \cline{2-5}
     & \texttt{R\_poi\_dis\_coord} & Estimate the distance between two POIs, with coordinates. & What is the approximate distance between \texttt{\{SI-SID$_0$\}} $(lon_0,lat_0)$ and \texttt{\{SI-SID$_1$\}} $(lon_1,lat_1)$? & 37.8~km \\
    \hline
    \hline
    \end{tabular}
    }
\end{table*}

\section{Details of the Spatial Cognition Evaluation Benchmark}
\label{app:mgcpt_bench}

To comprehensively evaluate the spatial cognition acquired by MG-CPT, we construct a dedicated benchmark for each industrial dataset, consisting of 18 tasks with 2{,}000 questions each. All questions are automatically generated from the corresponding city's POI, road, and navigation data, and are strictly disjoint from the MG-CPT training corpus. The benchmark follows the three-tier organization of MG-CPT: \texttt{B} (Basic-attribute) tasks query the intrinsic properties of a single POI, such as its coordinates or the area it belongs to; \texttt{R} (Relational) tasks examine pairwise spatial relations between POIs, covering distance (\texttt{dis}), direction (\texttt{dir}), and surroundings (\texttt{nearby}); and \texttt{S} (Systemic) tasks assess city-scale route understanding (\texttt{navi}), such as inferring the destination or the passed-by POIs of a navigation route. Each task is instantiated in a multiple-choice format, and six representative tasks are additionally instantiated in an open-ended format to test generation beyond option matching. Moreover, the \texttt{coord}/\texttt{no\_coord} suffix controls whether the coordinates are provided in the question: the former tests coordinate-grounded computation, whereas the latter removes all numeric hints and forces the model to rely solely on its internalized geographic knowledge. Table~\ref{tab:mgcpt_tasks} summarizes all tasks along with representative examples. In addition to the question and the ground-truth answer, each task is equipped with a task-specific system prompt, which assigns the model the role of a city geography expert and specifies the expected output format. The final evaluation input thus follows the template ``\texttt{[system prompt] + [question]}''. The system prompts for the two formats are instantiated as follows:

\begin{itemize}[leftmargin=1em]
    \item \textbf{Multiple-choice.} \textit{``You are a geography expert of {City}, familiar with its road network and POI locations. Please answer the question accurately. For multiple-choice questions (with options A/B/C/D), directly output the option letter.''}
    \item \textbf{Open-ended.} \textit{``You are a geography expert of {City}, familiar with its POI locations. Please answer the question accurately. For open-ended questions, directly output the answer (e.g., the direction, distance, or district).''}
\end{itemize}

\end{document}